\documentclass[sigconf]{acmart}
\usepackage{xcolor,soul} 
\usepackage{makecell}   
\usepackage{tcolorbox}  
\usepackage{multirow}
\usepackage{tcolorbox}  
\usepackage{array}  
\newcolumntype{?}{!{\vrule width 1pt}}

\copyrightyear{2021}
\acmYear{2021}
\setcopyright{acmcopyright}
\acmConference[ESEM '21]{ACM / IEEE International Symposium on Empirical Software Engineering and Measurement (ESEM)}{October 11--15, 2021}{Bari, Italy}
\acmBooktitle{ACM / IEEE International Symposium on Empirical Software Engineering and Measurement (ESEM) (ESEM '21), October 11--15, 2021, Bari, Italy}
\acmPrice{15.00}
\acmDOI{10.1145/3475716.3475781}
\acmISBN{978-1-4503-8665-4/21/10}

\begin{document}
\addtolength{\abovecaptionskip}{-1mm}
\addtolength{\belowcaptionskip}{-1mm}

\title{An Empirical Study of Rule-Based and Learning-Based Approaches for Static Application Security Testing}


\author{Roland Croft}
\affiliation{University of Adelaide}
\affiliation{Cyber Security Cooperative Research Centre}
\email{roland.croft@adelaide.edu.au}

\author{Dominic Newlands}
\affiliation{University of Adelaide}
\email{dominic.newlands@student.}
\email{adelaide.edu.au}

\author{Ziyu Chen}
\affiliation{Monash University}
\email{zche0071@student.}
\email{monash.edu}

\author{M. Ali Babar}
\affiliation{University of Adelaide}
\affiliation{Cyber Security Cooperative Research Centre}
\email{ali.babar@adelaide.edu.au}

\renewcommand{\shortauthors}{Croft et al.}

\begin{abstract}
\textbf{Background:} Static Application Security Testing (SAST) tools purport to assist developers in detecting security issues in source code. These tools typically use rule-based approaches to scan source code for security vulnerabilities. However, due to the significant shortcomings of these tools (i.e., high false positive rates), learning-based approaches for Software Vulnerability Prediction (SVP) are becoming a popular approach. 
\textbf{Aims:} Despite the similar objectives of these two approaches, their comparative value is unexplored. We provide an empirical analysis of SAST tools and SVP models, to identify their relative capabilities for source code security analysis. 
\textbf{Method:} We evaluate the detection and assessment performance of several common SAST tools and SVP models on a variety of vulnerability datasets. We further assess the viability and potential benefits of combining the two approaches. 
\textbf{Results:} SAST tools and SVP models provide similar detection capabilities, but SVP models exhibit better overall performance for both detection and assessment. Unification of the two approaches is difficult due to lacking synergies. 
\textbf{Conclusions:} Our study generates 12 main findings which provide insights into the capabilities and synergy of these two approaches. Through these observations we provide recommendations for use and improvement. 
\end{abstract}


\maketitle

\section{Introduction}
Software security is vital for organisations to avoid catastrophic exploits, but it is significantly more difficult to detect Software Vulnerabilities (SVs) than regular fault detection \cite{morrison2018}. Static Application Security Testing (SAST) tools purport to provide timely automated support for secure software development. SAST tools help developers to detect SVs during the coding phase, where it is relatively inexpensive to identify and fix security problems in source code. SAST tools, which primarily operate through rule-based approaches, statically examine source code for known vulnerable patterns that indicate the presence of potential SVs. Given SAST tools enable developers to quickly and cheaply perform quality assurance steps from a security perspective at an early stage of software development, these tools have gained sizeable traction, particularly in Open Source Software (OSS) communities \cite{beller2016}. However, SAST tools have had mixed success in making inroads in commercial software development practices as developers seem weary of their limitations, particularly the significant amounts of false positives \cite{christakis2016,johnson2013,le2021}. 

Recently, another source code security analysis approach, Software Vulnerability Prediction (SVP) \cite{hanif2021}, has been gaining significant attention for security assurance during implementation stage. An increasing number of research efforts are developing effective learning-based approaches for the identification of vulnerable code modules \cite{hanif2021}. These approaches use historic source code modules to devise data-driven approaches for detection or prediction of SVs \cite{coulter2020}. Rather than using pre-defined rules like SAST tools, SVP models automatically learn the rules and patterns for SV detection. 

Both of these approaches, SAST tools and SVP, are aimed at effectively and efficiently analyzing source code for SVs. However, there are several differences in the techniques of detecting security flaws in source code underpinning these two approaches. SAST tools provide granular warnings, but suffer limited individual coverage \cite{aloraini2019} and high rates of false positives \cite{imtiaz2019}; SVP models can produce effective SV classifications, but they are difficult to adopt due to demanding data requirements and need for data science expertise. Perhaps due to the disparity of these approaches, their use and analysis largely exists in isolation. 

Despite the increasing amount of literature on SAST tools and SVP approaches, there has been little empirical research on providing evidence-based comparison of these two types of security assurance approaches. Hence, we decided to empirically compare these two approaches to investigate their respective capabilities of source code security analysis during early stages of software development. We also aimed to empirically explore the potential viability and effect of combining the two approaches. 

An empirical study like ours ought to consider a variety of factors when selecting a SAST approach, such as escaped SVs (false negatives), wasted inspection efforts (false positives), and setup requirements. Different organizations will typically value these factors differently. For instance, a mission critical system would aim to ensure that all SVs are identified; whereas an agile startup would be more oriented towards speed and lower inspection efforts. As tool integration is expensive \cite{christakis2016,johnson2013}, most organisations would desire to adopt a singular solution. Hence, it is imperative that we outline the relative strengths and weaknesses of each approach in advance, to assist practitioners with their selection. 


To empirically determine the relative trade-offs of each approach, we selected and deployed a variety of publicly available SAST tools and replicable SVP models. We then evaluated their SV detection and assessment performance on several open-source datasets to determine the relative capabilities of each approach. 

Our empirical study is one of the first, if not the first, efforts aimed at contributing a large-scale assessment of both rule-based and learning-based approaches to identify their comparative capabilities for detecting SVs. The main findings of this study include: 
\begin{itemize}
    \addtolength\itemsep{1mm}
    \item Both approaches exhibit similar capabilities in terms of recall, but SVP models produce much better overall performance. 
    \item Both approaches are constrained in their capabilities for SV assessment, but most SAST tools are incapable of this task. 
    \item Effective unification of the two approaches is difficult as there is a lack of synergy. 
\end{itemize}

Our findings provide evidence-based insights for both developers and researchers. For developers, we inform the comparative value of these two approaches for source code security analysis; we also provide some recommendations that can be useful for their decisions about selecting and using one of these approaches. For researchers, we identify some necessary research opportunities based on the pain-points we have discovered for each approach. We discuss the details of our findings and their implications in section 5. Our datasets and scripts are publicly available from our reproduction package \cite{reproduction_package}. 

\section{Background and Related Work}
\subsection{Static Application Security Testing}
Static Application Security Testing (SAST) tools are defined as tools that can statically analyze source code or compiled versions to help identify potential security flaws \cite{OWASP_static}. SAST is commonly performed as part of code review during the implementation phase of a software development project \cite{shahriar2012}. These tools are desirable as they provide early and immediate security feedback, which is more efficient and cheaper than finding security flaws in source code at a later stage of softare development. 

SAST tools scan source code using a set of pre-defined security weaknesses (rules). Different tools use different techniques for scanning, such as pattern matching \cite{viega2002}, data-flow analysis \cite{kildall1973}, or symbolic execution \cite{xie2003}. The types of security warnings they can produce are often dependent on the used detection techniques. 

Researchers have conducted several user studies with software developers to identify major pain-points and areas of improvement \cite{christakis2016,johnson2013,oyetoyan2018}. These studies commonly identify that developers are reluctant to use or take appropriate actions on the SAST tools' outputs; these studies report the major reasons for lack of use of SAST tools as the excessive number of outputs, difficulty in customization, and lack of useful warning messages. To better understand the actual tool performance, several researchers have also performed bench-marking of selected SAST tools \cite{aloraini2019,diaz2013,kaur2020,wagner2014}. 

However, none of the existing works for analysis of SAST tools considers or evaluates learning based approaches. We aim to extend this analysis by additionally comparing SVP performance. Furthermore, our analysis and assessment criteria is more extensive than the existing bench-marking works, which have only evaluated the raw detection performance on singular data sources. 

\subsection{Software Vulnerability Prediction}
Software Vulnerability Prediction (SVP) is another approach to performing source code analysis to detect SVs or security risks early in software development. That is why SVP approaches have several characteristics similar to SAST \cite{coulter2020}. However, rather than searching for a set of predefined security weaknesses, SVP models aim to automatically learn SV knowledge and patterns from historical data. This process has seen continual technical advancement over the last decade through a large number of research efforts \cite{hanif2021}. 

The SVP process follows a standard pipeline. First, a model extracts data modules from historical software repositories, such as version control and bug tracking systems. These modules include both \emph{vulnerable} and \emph{clean} code, to help learn the distinction between the two classes. Informative features are then generated from the code, such as software metrics or code tokens, for a model to learn from using a specified classification method (e.g., random forests). The trained prediction model can then be used to classify whether or not incoming code modules are potentially vulnerable. 

There are two main approaches for SVP \cite{ghaffarian2017}: software metric based approaches, and vulnerable code pattern recognition. The former utilizes software metrics, such as code complexity or development characteristics, to help decide which code modules are at risk of containing SVs. This approach relies on the correlation of these metrics to the emergence of SVs, which has been reported in several studies \cite{chowdhury2011,shin2010}. However, this assumption often means that these approaches are inaccurate and unable to identify SVs explicity \cite{morrison2015}. The latter approach utilises the explicit code tokens to identify vulnerable patterns in source code. This approach has been shown to outperform software metrics \cite{walden2014}, and is the more popular approach in literature \cite{ghaffarian2017}.  

\subsection{Combined Approaches}
Whilst these two approaches largely exist in isolation, some studies have investigated their combination. 
Learning-based approaches have been used to enhance static analysis tools by reducing their false positives \cite{yoon2014}, or to assist with the output analysis by ranking tool warnings \cite{pereira2019,ribeiro2019}. 
Alternatively, the output of static analysis tools has also been used to enhance the capabilities of SVP models, but these attempts have produced uninspiring results \cite{gegick2007,rahman2014}. 

Rahman et al. \cite{rahman2014} performed a comparative study of static bug finders and statistical prediction approaches to identify synergistic aspects that may be leveraged by combining the two types of approaches. In terms of performance, they found that the static bug finders perform similar to the prediction models. They also reported that the output of the static tools did not improve the performance of statistical prediction techniques, but prediction models were able to produce better orderings of the static tool outputs than the natural ordering. Whilst our study has an overall goal that is similar to Rahman and colleagues' \cite{rahman2014} work of comparing learning-based method to traditional static analysis methods, there are several key differences that has resulted in unique findings from our study. 

Rahman et al. \cite{rahman2014} only examined software bugs, not vulnerabilities. SVs exhibit different characteristics to regular software bugs as they do not necessarily represent functional flaws \cite{shahriar2012}. Additionally, SVs are much more scarce \cite{shin2013} and harder to detect \cite{morrison2018}. As such, we also expect the characteristics and performance of source code security analysis approaches to differ substantially. SV mitigation also requires better understanding of code and potential consequences \cite{smyth2017}. Hence, we have also investigated assessment that is another vital task of SVs. Furthermore, Rahman et al. \cite{rahman2014} only provided a comparison of the inspection costs of each approach. We have instead focused on the performance and capabilities of each approach, as SV detection is much more critical; exploited vulnerabilities can lead to catastrophic consequences. 

\section{Research Method}
The goal of this research is to empirically investigate the comparative value of source code security analysis approaches. We aim to answer the following Research Questions (RQs): 

\begin{itemize}
    \addtolength\itemsep{2mm}
    \item \textbf{RQ1} \emph{What is the capability of SAST tools and SVP models for SV detection?} \\
    We first aim to identify the performance and efficacy of each source code security analysis approach. It is vital that approaches have practical capabilities for detecting SVs. 
    
    \item \textbf{RQ2} \emph{What is the capability of SAST tools and SVP models for SV assessment?} \\
    Another critical component of SV mitigation is assessment \cite{smyth2017}. Once we have detected a vulnerability, we must also be able to identify its type, so that we can infer potential impacts and appropriately plan mitigation. Hence, we seek to identify the capabilities of each approach for SV assessment. 
    
    \item \textbf{RQ3} \emph{Can these approaches complement each other?} \\
    Finally, we are interested in investigating the potential synergies between SAST tools and SVP models. Despite the differences of these two approaches, we aim to determine whether the similarity of the task can enhance their combined performance for a unified approach; the outputs of the SAST tools are used as inputs for an SVP model.
\end{itemize}

\begin{figure}[h]
  \centering
  \includegraphics[width=\linewidth]{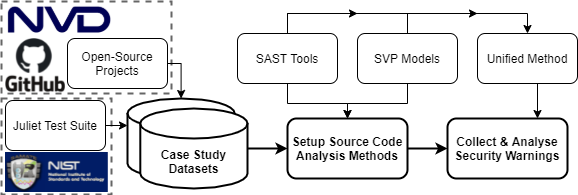}
  \caption{The overall methodology for evaluating SAST tools and SVP models.}
  \label{fig:method}
  \Description{A step by step workflow of how the data is extracted, tools/models are selected, and how the approaches are assessed.}
\end{figure}

To address these RQs, we have conducted a large-scale comparative study into the application of these approaches on a variety of SV datasets. Through this study, we have empirically evaluated the operation of a selected set of representative tools and SVP models. Figure \ref{fig:method} presents an overview of the study design. 

\subsection{Selecting and Extracting Case Study Datasets}
The goal of SAST tools is to help developers to identify SVs in source code. Hence, we aimed to collect a representative dataset of a codebase containing SVs. To make our set of SVs more generalizable to a real-world setting, we primarily constructed our datasets using real-world data from open-source repositories. 

Both of the approaches that we have investigated (SAST tools and SVP models) are language specific. It is common knowledge that vulnerable code patterns are difficult to translate across programming languages \cite{mushtaq2017,shahriar2012}. That is why we decided to collect the code datasets for C/C++. We made this decision as these two programming languages are generally known to have more SVs due to their lower levels of abstraction \cite{seacord2005}. C/C++ are also the 1st and 4th most popular programming languages respectively as per the March 2021 TIOBE index \cite{TIOBE}. 

For data collection from open source projects, we utilized the VulData7 framework developed by Jimenez et al. \cite{jimenez2018}. This framework automatically collects vulnerability report data (i.e., fixing commits of source code files) from the relevant software archives (Git and NVD reports) for open source projects. We formed our vulnerability dataset from pre-patch file versions. VulData7 is automatically configured to collect data for four repositories: Linux kernel, Wireshark, OpenSSL, and SystemD. However, we excluded SystemD due to the small initial size of this dataset (9 SVs). 

Additionally, to improve our data quality we attempted to address the limitations reported for the VulData7 tool \cite{jimenez2018} by performing two post processing steps. First, we removed all the duplicate data entries introduced from multiple code branches by dropping duplicate file contents. Second, we ensured that the vulnerability reports were actually relevant by removing fixing commits which do not make functional changes to code (i.e., only changes to comments or white space). Three of the authors manually validated a random sample (n=67, significant size, \cite{cochran2007}) of entries to ensure that the quality of our data is sufficient (that they contain a relevant vulnerability to the associated security report). Disagreements were resolved through discussions. We found that 90\% of our entries were valid at a confidence level of 90\% +/-10\%. 

We also required examples of non-vulnerable files so that we can train SVP models and report false positives. Hence, we obtained a set of C/C++ files which were not labelled as vulnerable to form our non-vulnerable class. As our vulnerable files come from different versions of each repository, we similarly sampled non-vulnerable files from different commits made to each repository. To help ensure that this assumption of non-vulnerability was accurate, we only considered files which did not have any associated vulnerability reports across all versions. This is because we were unable to accurately determine in which version a vulnerability is introduced or removed in a file \cite{fan2020}; fixing commits may only be partial or incomplete. Additionally, we excluded the file versions where the commit message contained a security keyword (using the security keyword list from Le et al. \cite{le2020}) as these files may be related to SVs. Finally, we only kept unique files to avoid redundancy.

Given the open-source repositories use security advisories to record SV data, they are only representative of SVs detected during the testing or maintenance phases. To this extent, these data sources do not document SVs that have been removed during implementation or code review \cite{paul2021}, which is a conundrum as it is these very SVs that SAST tools are oriented towards \cite{OWASP_static}. Hence, we complemented our dataset with the inclusion of a synthetic test suite to capture more conventional SVs. We selected the Software Assurance and Reference Dataset (SARD) produced by the National Institute of Standards and Technology (NIST) \cite{SARD}. We downloaded all the test cases from version 1.3 of the Juliet Test Suite for C/C++. This test suite provides labelled examples of both clean and vulnerable code. Hence, to obtain the file examples of both classes, we split the test cases into separate vulnerable and non-vulnerable files. Table \ref{tab:data_stats} reports the summary statistics of our collected datasets. 

\begin{table}[h]
  \caption{Statistics of the selected datasets.}
  \label{tab:data_stats}
  \resizebox{\columnwidth}{!}{%
  \begin{tabular}{cccccc}
    \hline
    \makecell{\textbf{Data Source}} & \makecell{\textbf{Project}} & \makecell{\textbf{\% Vulner-}\\\textbf{able Files}} & \makecell{\textbf{Total}\\\textbf{Files}} & \textbf{\# LOC} & \makecell{\textbf{\# CWE}\\\textbf{Types}}\\
    \hline
    \multirow{3}{*}{\makecell{Open-Source\\Projects}} & OpenSSL & 32.37\% & 2258 & 1,567,897 & 15\\
    & Wireshark & 9.06\% & 3731 & 5,678,133 & 12\\
    & Linux Kernel & 7.66\% & 34,961 & 25,600,912 & 18\\
    \hline
    \makecell{Synthetic\\Test Cases} & \makecell{Juliet Test\\Suite} & 49.61\% & 200,159 & 16,324,088 & 13\\
    \hline
\end{tabular}%
}
\end{table}

\subsection{Selecting SAST Tools}
For SAST tool selection, we first analysed the list of tools documented by NIST \cite{NIST_tools} and OWASP \cite{OWASP_static}. We applied two restrictions when selecting the SAST tools for this study. First, the tools must be open source as we aimed to consider the most accessible and widely used tools. Second, we did not consider tools which had usage limits or purely operate through a graphical user interface, as we aimed to conduct a large scale benchmarking of these tools. 
Based on these requirements, we selected 3 tools: Flawfinder, Cppcheck, and RATS. These tools have been commonly used in prior works \cite{aloraini2019,kaur2020} or by large organizations \cite{christakis2016}. We provide a brief description of each tool:  

\emph{Flawfinder} \cite{flawfinder} uses a simple pattern matching method to match source code text with a built-in database of known vulnerable C/C++ functions. The tool is designed to be fast, rather than accurate. 

\emph{Cppcheck} \cite{cppcheck} uses a unique bi-directional data flow analysis method to focus on detecting SVs resulting from undefined behaviour. Its goal is to produce very few false positives. Cppcheck additionally provides style and code quality warnings, but these SVs are ignored for our study as they are not included in our datasets. 

\emph{RATS} \cite{rats} is a Rough Auditing Tool for Security, which uses a simple pattern matching method similar to Flawfinder. Hence, it is also oriented towards speed rather than performance.

We represented the collected vulnerable and non-vulnerable files as a single codebase for each of the four datasets, which we fed as input to the three selected SAST tools\footnote{For Juliet, we excluded warnings relating to \emph{srand}, as these are located in the main function of Juliet test cases, making them consistent for all test cases.}. 

\subsection{Building SVP Models}
There are two major approaches for SVP models, determined by the utilised features: software metrics and code tokens. We build SVP models for these two approaches by using the features proposed by Munaiah and Meneely \cite{munaiah2019}, and Scandariato et al. \cite{scandariato2014}. These models operate at the file-level; the most common granularity of existing SVP techniques \cite{ghaffarian2017}. 

Munaiah and Meneely \cite{munaiah2019} proposed and evaluated 10 software metrics for SVP. For our replication, we used eight of the software metrics\footnote{The used software metrics are available in our reproduction package \cite{reproduction_package}.}. We excluded the \emph{\# Paths} metric as we found it had a high correlation to cyclomatic complexity, and the \emph{Offender} metric as it aligned with the rules of our class separation. The \emph{Contribution, Collaboration} and \emph{Churn} metrics could not be extracted for the Juliet dataset. Although a higher number of software metrics have been used in previous studies \cite{zimmermann2010}, these metrics often overlap or correlate, which can hinder the performance of a model \cite{ghotra2017}.  

For code tokens, we used text mining to extract Bag-of-Word features as proposed by Scandariato et al. \cite{scandariato2014}. We built a custom tokenizer for C/C++ to appropriately handle code syntax, which is available in our reproduction package \cite{reproduction_package}. For our replication, we excluded the comments from tokenization and limited the vocabulary size to 1000 tokens. For the Juliet dataset, we replaced function names with neutral strings as we found the original function names to be separated between the \textit{clean} and \textit{vulnerable} files. 

We additionally created a combined SVP model using both software metric and code token features, for the purpose of direct comparison with SAST tools. However, the performance of the combined model did not differ to just using code token features, so we do not report its performance separately. The models were trained using a variety of common machine learning classification algorithms \cite{yang2020}: K-Nearest Neighbors, Support Vector Machine, Random Forest, and AdaBoost. We observed that the Random Forest classifier produced the best performance values for each experiment, except for the Juliet dataset when using code token features, and hence we refer to this algorithm the most. 

Additionally, to address RQ3 we experimented with creating a unified method; an SVP model using features created from the output of SAST tools. The SAST tool features were inspired by the features proposed by Ribeiro et al. \cite{ribeiro2019} for ranking SAST tool output. For each file, we considered the number of warnings and mean severity level for each tool. Additionally, we considered the aggregated features for the three tools as we expected multiple warnings to be more likely to represent a true vulnerability. We used the mean number of warnings and normalized severity of the three SAST tools, the ratio of warnings that overlap, and the average number of warnings within four lines of another warning.   

To properly construct and evaluate our SVP models, the appropriate hyperparameters\footnote{The tested hyperparameters and values are listed in our reproduction package \cite{reproduction_package}.} of each model were tuned using cross-validated grid search on the training set. To evaluate the performance of our models, we followed the recommendations by Tantithamthavorn et al. \cite{tantithamthavorn2016} to use out-of-sample bootstrap for evaluation. Out-of-sample bootstrap constructs a training set using random sampling with replacement from the original dataset. The model was then trained and validated on the random training set, and finally evaluated using the remaining unseen entries of the dataset which were not sampled. Although time-based validation is a recommended evaluation process as it better represents the real world scenario \cite{falessi2020}, it can produce performance estimates that are often biased or unstable \cite{tantithamthavorn2016}. Hence, we used out-sample-bootstrap to obtain more stable performance evaluations for the purposes of our comparative analysis. Due to the randomness of this process, we averaged the performance of 30 runs.

\subsection{Evaluating Tool/Model Outputs}
\begin{table*}[t]
  \caption{Performance comparison of the source code security analysis approaches.}
  \label{tab:performance}
  \resizebox{\textwidth}{!}{%
  \begin{tabular}{|c||c|c|c|c|c|c|c|c|c|c|c|c|c|c|c|c|c|c|c|c|c|c|c|c|}

    \multicolumn{25}{c}{\Large\textbf{SAST Tools}}\\
    \hline
    \multicolumn{1}{|c|?}{\multirow{2}{*}{\textbf{Project}}} & \multicolumn{8}{c?}{\textbf{Flawfinder}} & \multicolumn{8}{c?}{\textbf{Cppcheck}} & \multicolumn{8}{c?}{\textbf{RATS}}\\
    \cline{2-25}
    \multicolumn{1}{|c|?}{} & \multicolumn{2}{c|}{\# Warnings} & \multicolumn{2}{c|}{Recall} & \multicolumn{2}{c|}{Precision} & \multicolumn{2}{c?}{MCC} & \multicolumn{2}{c|}{\# Warnings} & \multicolumn{2}{c|}{Recall} & \multicolumn{2}{c|}{Precision} & \multicolumn{2}{c?}{MCC} & \multicolumn{2}{c|}{\# Warnings} & \multicolumn{2}{c|}{Recall} & \multicolumn{2}{c|}{Precision} & \multicolumn{2}{c?}{MCC}\\
    \hline
    \hline
   
    \multicolumn{1}{|c|?}{OpenSSL} & \multicolumn{2}{c|}{52.08\%} & \multicolumn{2}{c|}{0.732} & \multicolumn{2}{c|}{0.455} & \multicolumn{2}{c?}{0.292} & \multicolumn{2}{c|}{30.96\%} & \multicolumn{2}{c|}{0.703} & \multicolumn{2}{c|}{0.735} & \multicolumn{2}{c?}{0.589} & \multicolumn{2}{c|}{55.67\%} & \multicolumn{2}{c|}{0.744} & \multicolumn{2}{c|}{0.433} & \multicolumn{2}{c?}{0.261}\\
    \hline
    \multicolumn{1}{|c|?}{Wireshark} & \multicolumn{2}{c|}{29.21\%} & \multicolumn{2}{c|}{0.609} & \multicolumn{2}{c|}{0.189} & \multicolumn{2}{c?}{0.220} & \multicolumn{2}{c|}{14.90\%} & \multicolumn{2}{c|}{0.533} & \multicolumn{2}{c|}{0.324} & \multicolumn{2}{c?}{0.340} & \multicolumn{2}{c|}{27.10\%} & \multicolumn{2}{c|}{0.571} & \multicolumn{2}{c|}{0.191} & \multicolumn{2}{c?}{0.213}\\
    \hline
    \multicolumn{1}{|c|?}{Linux} & \multicolumn{2}{c|}{37.57\%} & \multicolumn{2}{c|}{0.676} & \multicolumn{2}{c|}{0.138} & \multicolumn{2}{c?}{0.178} & \multicolumn{2}{c|}{25.87\%} & \multicolumn{2}{c|}{0.710} & \multicolumn{2}{c|}{0.210} & \multicolumn{2}{c?}{0.297} & \multicolumn{2}{c|}{3.55\%} & \multicolumn{2}{c|}{0.195} & \multicolumn{2}{c|}{0.421} & \multicolumn{2}{c?}{0.249}\\
    \hline
    \multicolumn{1}{|c|?}{Juliet} & \multicolumn{2}{c|}{47.71\%} & \multicolumn{2}{c|}{0.488} & \multicolumn{2}{c|}{0.507} & \multicolumn{2}{c?}{0.021} & \multicolumn{2}{c|}{5.80\%} & \multicolumn{2}{c|}{0.075} & \multicolumn{2}{c|}{0.638} & \multicolumn{2}{c?}{0.071} & \multicolumn{2}{c|}{11.83\%} & \multicolumn{2}{c|}{0.128} & \multicolumn{2}{c|}{0.535} & \multicolumn{2}{c?}{0.029}\\
    \hline
   
    \multicolumn{25}{c}{}\\
    
    \multicolumn{25}{c}{\Large\textbf{SVP Models}}\\
    \hline
    \multicolumn{1}{|c|?}{\multirow{2}{*}{\textbf{Project}}} & \multicolumn{12}{c?}{\textbf{Software Metrics}} & \multicolumn{12}{c?}{\textbf{Code Tokens}}\\
    \cline{2-25}
    \multicolumn{1}{|c|?}{} & \multicolumn{3}{c|}{\# Warnings} & \multicolumn{3}{c|}{Recall} & \multicolumn{3}{c|}{Precision} & \multicolumn{3}{c?}{MCC} & \multicolumn{3}{c|}{\# Warnings} & \multicolumn{3}{c|}{Recall} & \multicolumn{3}{c|}{Precision} & \multicolumn{3}{c?}{MCC}\\
    \hline
    \hline
    
    \multicolumn{1}{|c|?}{OpenSSL} & \multicolumn{3}{c|}{30.59\%} & \multicolumn{3}{c|}{0.845} & \multicolumn{3}{c|}{0.892} & \multicolumn{3}{c?}{0.795} & \multicolumn{3}{c|}{30.71\%} & \multicolumn{3}{c|}{0.915} & \multicolumn{3}{c|}{0.963} & \multicolumn{3}{c?}{0.904}\\
    \hline
    \multicolumn{1}{|c|?}{Wireshark} & \multicolumn{3}{c|}{6.27\%} & \multicolumn{3}{c|}{0.495} & \multicolumn{3}{c|}{0.721} & \multicolumn{3}{c?}{0.561} & \multicolumn{3}{c|}{5.90\%} & \multicolumn{3}{c|}{0.554} & \multicolumn{3}{c|}{0.856} & \multicolumn{3}{c?}{0.660}\\
    \hline
    \multicolumn{1}{|c|?}{Linux} & \multicolumn{3}{c|}{3.67\%} & \multicolumn{3}{c|}{0.413} & \multicolumn{3}{c|}{0.863} & \multicolumn{3}{c?}{0.565} & \multicolumn{3}{c|}{4.66\%} & \multicolumn{3}{c|}{0.565} & \multicolumn{3}{c|}{0.930} & \multicolumn{3}{c?}{0.709}\\
    \hline
    \multicolumn{1}{|c|?}{Juliet} & \multicolumn{3}{c|}{52.60\%} & \multicolumn{3}{c|}{0.893} & \multicolumn{3}{c|}{0.843} & \multicolumn{3}{c?}{0.729} & \multicolumn{3}{c|}{53.14\%} & \multicolumn{3}{c|}{0.922} & \multicolumn{3}{c|}{0.862} & \multicolumn{3}{c?}{0.778}\\
    \hline
    \multicolumn{25}{p{\textwidth}}{\textbf{Note: The \# Warnings column shows the percentage of files flagged. For the SVP models, all values are averaged over 30 runs.}}\\
\end{tabular}%
}
\end{table*}

To form a comparison of SAST tools and SVP models, we needed to evaluate both on an equal footing, i.e., at the same level of granularity. To achieve this, we considered the warnings of both approaches at the file-level; whether a file was flagged as vulnerable or not. Whilst SAST tools produce individual warnings at the line-level, SV mitigation does not operate at this granularity in reality. To properly identify a vulnerability, a developer must inspect and understand a much larger code context \cite{smith2018}. For our analysis, we expanded this context to match that of SVP models at the file-level. As SAST tools can produce multiple warnings for a single file, we aggregated the SAST tool outputs into a single classification per file. True positives are vulnerable files with warnings, false positives are non-vulnerable files with warnings, false negatives are vulnerable files without warnings, and true negatives are non-vulnerable files without warnings. 

We evaluated the approaches for two main tasks; detection and assessment. Ultimately, the goal of source code security analysis is to identify potential SVs in source code. Hence, detection is the most important capability of each approach. However, unlike other types of code defects that may cause software to function unexpectedly or incorrectly, SVs expose code to potential exploits. Hence, we also need to assess the SV type so that we can better understand the relevant nature and impacts of SVs. 

For the assessment metric, we used the Common Weakness Enumeration (CWE) \cite{CWE} to classify SV type. Through the CWE type, we can understand the potential severity and impacts. As the labels in our dataset come from NVD or synthetic cases, the majority of vulnerable entries have an associated CWE type to serve as a ground truth. CWE uses a hierarchical structure, so we grouped CWEs to their highest level category to lower the dimensionality of our analysis, similar to Paul et al. \cite{paul2021}. For SAST tools, we considered whether they were able to produce a warning of the correct type for a file. For SVP models, we evaluated their ability for assessment through multi-class SVP models, which predict the CWE type (or no type for non-vulnerable cases) of a file. 

To measure the effectiveness and assessment, we used the performance measures of precision, recall, and Matthew's Correlation Coefficient (MCC). Although the F1 Score is often used as an overall indicator of model performance, MCC provides a more reliable statistic which considers all four confusion matrix categories \cite{chicco2020}.  Hence, we considered MCC for selecting optimal SVP models and parameters. Precision and recall have a range from 0 to 1, while MCC has a range of –1 to 1, where 1 is the best value for all metrics. 

We also conducted additional manual inspection of the output of SAST tools and SVP models on a random 10\% sample of files for each dataset (using the remaining 90\% of files as training data for SVP models), to better determine the detection capabilities of each of the compared approaches. We refer to this as our comparison set. We manually analysed up to 15 random files that SAST tools and SVP models each flagged as true positives, false positives and false negatives, exclusively and in combination. In total, we manually analysed the classifications of 265 files. 

\section{Results and Analysis}
\subsection{RQ1: What is the capability of SAST tools and SVP models for SV detection?}
Table \ref{tab:performance} displays the performance metrics of each of the studied approaches on the four projects' datasets. In terms of precision and MCC, each of the SVP models outperformed all three of the SAST tools (0.54 higher precision and MCC on average across the four datasets). We confirmed the performance increase to be significant using the Wilcoxon signed rank test \cite{wilcoxon1992} on the performance difference for SAST tools compared to SVP models for each of the four datasets (p = 0.002 for precision and MCC scores). However, the recall rate of SAST tools was comparable to, and in some instances even better than that of SVP models' recall. Likewise, using the Wilcoxon signed rank test \cite{wilcoxon1992}, we failed to reject the null hypothesis that recall rate is different between SAST tools and SVP models (p = 0.1). However, the high recall rate of SAST tools came at a significant trade-off to precision due to the high number of false positives that these tools produced. 

\begin{tcolorbox}[right=1pt, left=1pt, top=1pt, bottom=1pt, colback=white]
    \textbf{Finding 1:} \textit{We cannot conclude that SAST tools and SVP models produce different recall values.}
\end{tcolorbox}

\begin{tcolorbox}[right=1pt, left=1pt, top=1pt, bottom=1pt, colback=white]
    \textbf{Finding 2:} \textit{SVP Models exhibit significantly better precision and overall performance.}
\end{tcolorbox}

\begin{table}[h]
  \caption{Jaccard similiarity coefficient of SAST tool and SVP model warnings for vulnerable files.}
  \label{tab:similarity}
  \begin{tabular}{ccccc|c}
     & \textbf{OpenSSL} & \textbf{Wireshark} & \textbf{Linux} & \textbf{Juliet} & \textbf{Mean}\\
    \hline
    \textbf{Similarity} & 0.86 & 0.43 & 0.61 & 0.53 & 0.61\\
\end{tabular}
\end{table}

We examined the similarity of predictions for the vulnerable files in our comparison set for the aggregated SAST tool outputs and the combined SVP model. Table \ref{tab:similarity} displays the Jaccard similarity coefficient \cite{jaccard1912}. We observed that the detected vulnerabilities for SAST tools and SVP models were overall 61\% similar across the four projects. Considering that the recall value for these two approaches was also not significantly different, we claim that the detection capabilities of these two approaches are similar. 

\begin{tcolorbox}[right=1pt, left=1pt, top=1pt, bottom=1pt, colback=white]
    \textbf{Finding 3:} \textit{The detection capabilities of the positive class are similar for SAST tools and SVP approaches.}
\end{tcolorbox}

Unexpectedly, SAST tools produced the lowest performance (in terms of recall and MCC) on the Juliet dataset. An MCC value that is close to 0 indicates that the approach is doing little better than random guessing. We expected SAST tools to perform better on this dataset as it is of lower complexity; SVs are more distinct and rigid. However, upon manual inspection we discovered the difficulties to stem from the small class separation in this dataset; the vulnerable and non-vulnerable files only differentiate by minor line changes that alter the security. SAST tools were unable to effectively differentiate between vulnerable and non-vulnerable files; they often flagged both files or none of them. This reinforces the notion that SAST tools produce extremely large numbers of false positives that makes it difficult to identify true vulnerabilities \cite{christakis2016,johnson2013}. However, SVP models were able to achieve good performance values on both the open-source datasets and the Juliet dataset. This shows that they were able to differentiate between minor changes in code as well as detect SVs of a varying complexity. 

For SVP models, software metrics performed worse than code tokens. SVP models also generally performed worse for the Wireshark and Linux datasets. This is likely because these datasets suffer from class imbalance issues \cite{tantithamthavorn2018}. 

Across the three SAST tools, Flawfinder was able to produce the highest recall values. However, Flawfinder also produced the highest number of warnings across the four projects. Contrastingly, Cppcheck achieved the highest precision but produced the lowest number of warnings across the four projects. However, the recall of Cppcheck was not higher than the other two SAST tools, which indicates a trade-off between recall and precision in these tools. Tools which flag abundantly are more likely to identify both true and false positives. It is very difficult for these tools to achieve both high precision and recall.

\begin{tcolorbox}[right=1pt, left=1pt, top=1pt, bottom=1pt, colback=white]
    \textbf{Finding 4:} \textit{SAST tools exhibit a trade-off between precision and recall, based on the number of warnings.}
\end{tcolorbox}

SVP models generally flagged a lower percentage of files to SAST tools. The flagged number of files was more representative of the true distribution of vulnerable files seen in Table \ref{tab:data_stats}. Hence, these models achieved much higher values of precision compared to SAST tools. Despite this disparity, we would actually expect the inspection costs of SAST tools to be much lower due to the finer granularity of the line-specific warnings they provide. However, upon manual inspection we found this not to be the case as the line-level warnings were extremely inaccurate. For the vulnerable files of the open source datasets that SAST tools flagged, we found that only 5\% had an appropriate line-level warning. 

\begin{tcolorbox}[right=1pt, left=1pt, top=1pt, bottom=1pt, colback=white]
    \textbf{Finding 5:} \textit{SVP models produce a lower number of files to inspect.}
\end{tcolorbox}

\begin{table}[h]
  \caption{Recall of line-level predictions for SAST tools.}
  \label{tab:line_level}
  \begin{tabular}{cccc}
    \hline
    \textbf{Project} & \textbf{Flawfinder} & \textbf{Cppcheck} & \textbf{RATS}\\
    \hline
    OpenSSL & 0.009 & 0 & 0.016\\
    Wireshark & 0.011 & 0.001 & 0.008\\
    Linux & 0.008 & 0.002 & 0.004\\
    Juliet & 0.167 & 0.036 & 0.057\\
    \hline
\end{tabular}
\end{table}

Table \ref{tab:line_level} displays the recall of SAST tools at the line-level. We identified vulnerable lines through the vulnerability fixing commit changes for the open source datasets, and through the Juliet meta-data. All three tools struggled to produce any true positive warnings for any of the projects. We note that the line warnings were more useful for the Juliet dataset; from our manual inspection, we found that 77\% of the true positives had valid line warnings. 

We found that most of the SAST tool warnings were incidental; unrelated to the true nature of a vulnerability. For Flawfinder and RATS, the warnings predominantly stemmed from the use of string handling functions, e.g., \emph{memcpy, char} and \emph{strlen}. For Cppcheck, the bulk of errors came from syntax issues. For the open source datasets, however, the true vulnerabilities typically occurred through missing logic and checks, i.e., checking for the length of a buffer or existence of a pointer. As the vulnerable code was often contained in for/if statements, rather than function use, it was very difficult for SAST tools to detect or localize these.  

\begin{tcolorbox}[right=1pt, left=1pt, top=1pt, bottom=1pt, colback=white]
    \textbf{Finding 6:} \textit{The majority of SAST tool warnings are incidental.}
\end{tcolorbox}

Despite these incidental warnings, SAST tools were still able to produce warnings at the file-level with high recall. We suspect this was because files containing vulnerabilities tend to be more complex \cite{munaiah2019}. Hence, they also tend to use more approaches  flagged by SAST tools. This correlation is unlikely to be strong due to the low precision of these methods. However, in this sense SAST tool warnings operate similarly to software metrics as indicators of vulnerabilities. We investigated this correlation further in RQ3. 

SAST tools were also unable to predict vulnerabilities stemming purely from methods calls; functions defined externally in separate files. This is because SAST tools are unable to infer any information about the surrounding semantics of these calls. SVP models were able to correctly classify some of these limited context files as vulnerable as they analyse the file as a whole. However, they similarly struggled in their classification as they produced many false positives and negatives. 

In continuation of our manual analysis, we found it was very difficult to discern between what SVP models could and could not predict. This is because they made predictions at the whole file level, whilst the actual vulnerable lines were usually small and consistent. For instance, the vulnerable code may be the same for multiple files but SVP models produced different positive and negative predictions for each file. The predictions were not reliable; and there was no way of interpreting why a model had made a particular prediction.

\begin{tcolorbox}[right=1pt, left=1pt, top=1pt, bottom=1pt, colback=white]
    \textbf{Finding 7:} \textit{The prediction capabilities of SVP models are not transparent.}
\end{tcolorbox}

\begin{figure}[h]
  \centering
  \includegraphics[width=\linewidth]{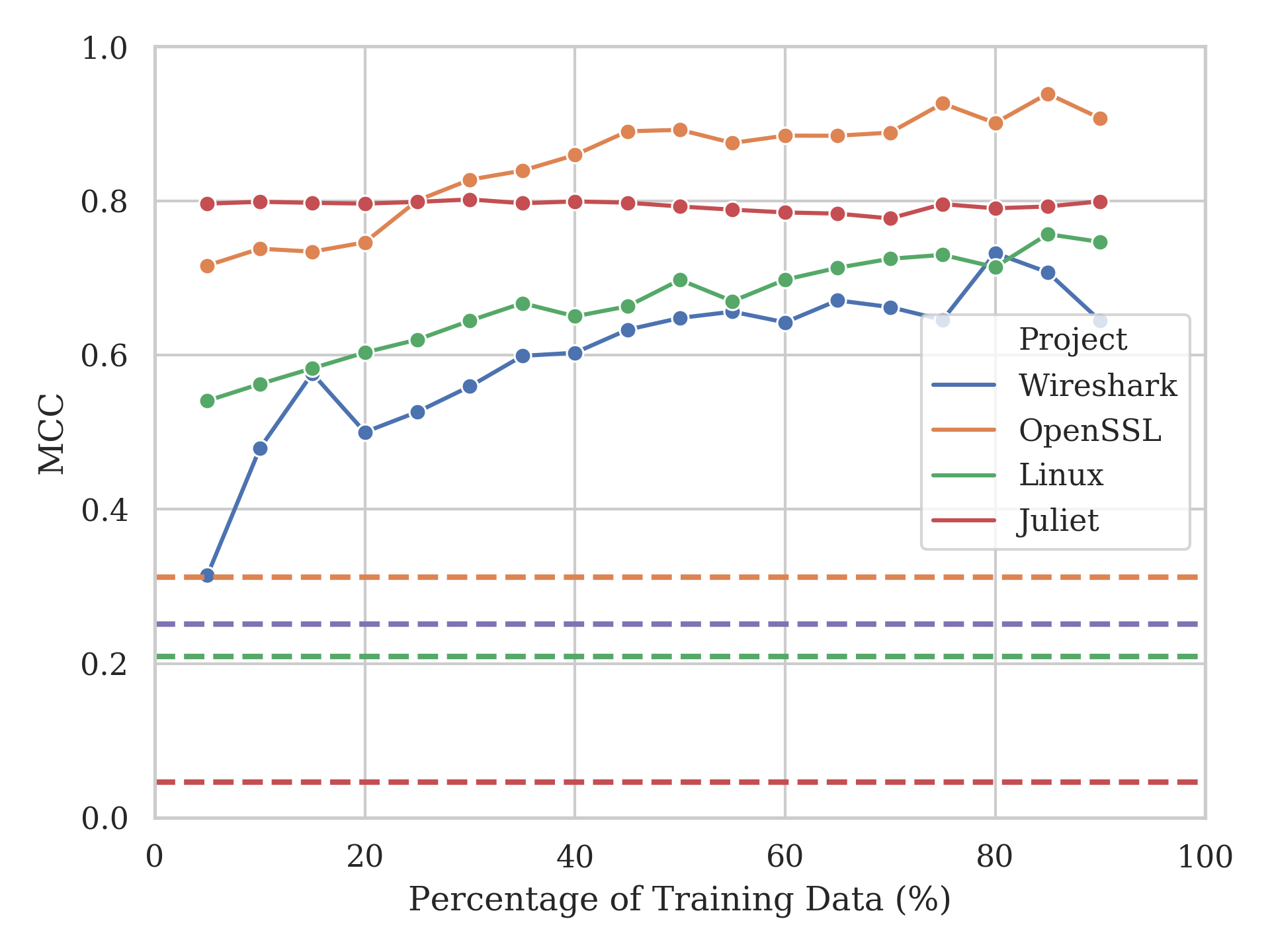}
  \caption{Model Performance for Increasing Percentages of Training Data. Note: MCC of the SAST tools is marked via a dashed line}
  \label{fig:data_impact}
  \Description{A line plot with MCC on the y-axis and training data percentage on the x-axis for the four datasets. Note: MCC of the SAST tools are presented with a dashed line. }
\end{figure}

It should be noted that SVP models were only evaluated on a portion of the data, whereas SAST tools were applied to the entire dataset. This is due to the large training requirements and data hungriness of SVP models. Hence, we also examined the impacts of data requirements on SVP models, i.e., how much data an organisation actually requires in order to use these models. Using the best SVP model configuration, we measured the performance impacts of reducing the amount of training data in increments of 5\%. Figure \ref{fig:data_impact} displays the MCC value for each model for varying percentages of training data, as well as the MCC of using SAST tools in combination. 

SVP model performance was still greater than SAST tools even when using as little as 5\% of the available data. Hence, early application of these models is not unreasonable, and efforts should be made to implement models as soon as data is made available. However, as SAST tools do not have data requirements, these tools can even be used at project inception, potentially to assist in acquiring training data for SVP models. 

Using the Kendall rank correlation coefficient \cite{kendall1938}, we observed a positive correlation between the amount of training data and the model performance for each of the open source datasets (p \textless 0.01), but not the Juliet dataset. This increasing performance shows that SVP models continue to improve with the maturity of a system, as new data becomes available. 

\begin{tcolorbox}[right=1pt, left=1pt, top=1pt, bottom=1pt, colback=white]
    \textbf{Finding 8:} \textit{SAST tools can be used by a project immediately. However, SVP models exhibit better performance than SAST tools even with low data requirements.}
\end{tcolorbox}

\subsection{RQ2: What is the capability of SAST tools and SVP models for SV assessment?}
Vulnerability assessment is a vital task to assist with SV mitigation \cite{khan2018}.  Table \ref{tab:assessment} reports the assessment performance of each approach. 

\begin{table}[h]
  \caption{MCC of the assessment task for source code security analysis approaches. Note: RATS does not produce assessment information.}
  \label{tab:assessment}
  \resizebox{\columnwidth}{!}{%
  \begin{tabular}{|c?cc?cc?}
    \cline{2-5}
    \multicolumn{1}{c?}{} & \multicolumn{2}{c?}{\textbf{SAST Tools}} & \multicolumn{2}{c?}{\textbf{SVP Models}}\\
    \hline
    \textbf{Project} & \textbf{Flawfinder} & \textbf{Cppcheck} & \makecell{\textbf{Software}\\\textbf{Metrics}} & \makecell{\textbf{Code}\\\textbf{Tokens}}\\
    \hline
    OpenSSL & 0.208 & -0.033 & 0.583 & 0.661\\
    Wireshark & 0.215 & -0.012 & 0.359 & 0.435\\
    Linux & 0.156 & -0.022 & 0.346 & 0.437\\
    Juliet & 0.235 & 0.102 & 0.575 & 0.769\\
    \hline
\end{tabular}%
}
\end{table}

SVP models again performed well for the assessment task, but experienced a performance decrease. This is to be expected as the class imbalance is further exacerbated in the multi-class problem, making assessment a more difficult task for learning-based approaches.

However, SAST tools were largely unable to effectively perform assessment; only Flawfinder produced the values which were a little better than random guessing. This incapability could be due to a few reasons: i) as discussed in section 4.1, the majority of the SAST tool warnings for the open source datasets were incidental; hence, it did not reflect the true nature of the vulnerability. This would be similarly perpetuated for SV assessment; ii) due to the rule-based nature of the SAST tool approach, the tools are only setup to detect certain types of SVs, which does not cover all the SVs present in a real-world scenario. This would similarly explain the poor performance of SAST tools on the Juliet dataset, as this dataset contains a wide array of different SV types; iii) assessment is not the major objective of tools, i.e., RATS does not even provide information for this task. Whilst the tools will often provide their own form of classification for their output (e.g., \textit{fixed global buffer size}), these do little to assist with assessment and understanding of the impacts. 

\begin{tcolorbox}[right=1pt, left=1pt, top=1pt, bottom=1pt, colback=white]
    \textbf{Finding 9:} \textit{Assessment performance is worse than detection performance for both approaches.}
\end{tcolorbox}

\begin{tcolorbox}[right=1pt, left=1pt, top=1pt, bottom=1pt, colback=white]
    \textbf{Finding 10:} \textit{SAST tools are generally incapable of assessment.}
\end{tcolorbox}

SVP models have a similar prediction constraints as they are limited to the contents of the training data. They can only provide accurate classifications for CWE types that have regularly occurred in the past. Hence, models struggle to predict SV types with few examples, which is exacerbated by the scarcity and imbalance of SV data, and are completely incapable of predicting unseen types. 

The presence of these limitations was reinforced through our manual inspection, as we observed both SVP models and SAST tools struggled to flag more obscure vulnerabilities. False negatives often came from vulnerabilities that were not as well represented in the data, such as race conditions and timing attacks. 

\begin{tcolorbox}[right=1pt, left=1pt, top=1pt, bottom=1pt, colback=white]
    \textbf{Finding 11:} \textit{Both approaches are constrained in the types of vulnerabilities that they can assess.}
\end{tcolorbox}

\subsection{RQ3: Can these approaches complement each other?}
To evaluate the combined performance of these two approaches, we considered two approaches for unification: a naive approach in which we simply merge the outputs of the two approaches, and a more sophisticated approach in which we use the outputs of SAST tools as features for SVP models (as described in Section 3.3). 

\begin{figure}[h]
  \centering
  \includegraphics[width=\linewidth]{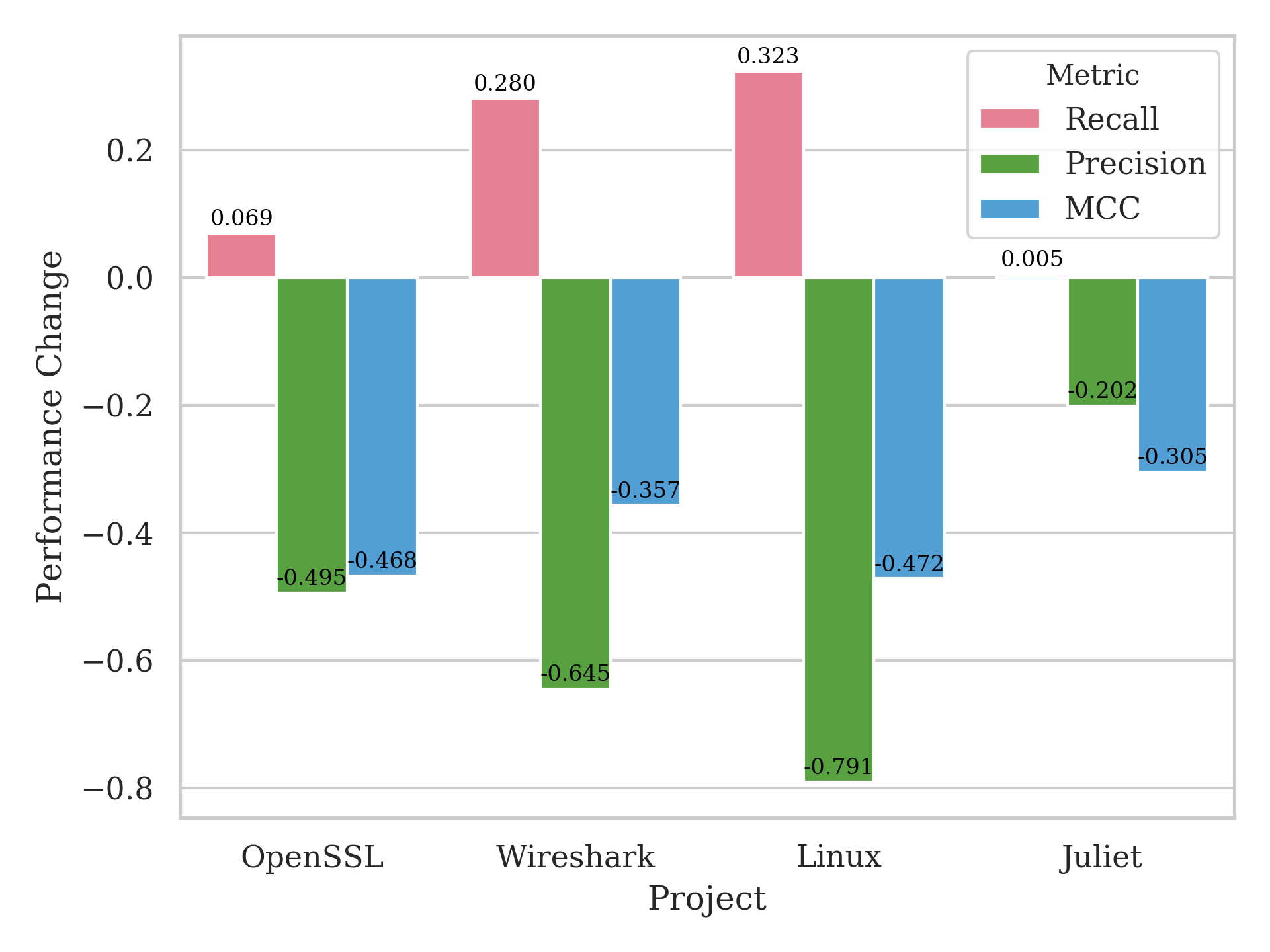}
  \caption{Difference in performance for merging SAST tool and SVP model outputs in comparison to just using SVP models}
  \label{fig:sast_svp_chain}
  \Description{A bar chart with Performance change on the y-axis and project on the x-axis for the three performance metrics recall, precision and mcc.}
\end{figure}

We first investigated the simple approach of using both approaches in parallel. We considered the merged outputs of all three SAST tools and the combined SVP model to identify flagged files. Figure \ref{fig:sast_svp_chain} displays the mean performance change in each dataset in comparison to the original SVP model performance. 

By merging the outputs from both approaches, the recall of these approaches improved significantly (except for Juliet due to the low recall for SAST tools on this dataset). For the Wireshark and Linux dataset, there were a high number of false negatives for SVP models, which SAST tools were able to flag. Hence, if detecting as many vulnerabilities as possible is the sole objective of source code security analysis regardless of the inspection efforts, then using multiple approaches is a good approach. However, the increase in recall came at a significant trade-off to precision due to the general imprecision of SAST tools. The use of multiple approaches results in an overall performance decrease. 

\begin{figure}[h]
  \centering
  \includegraphics[width=\linewidth]{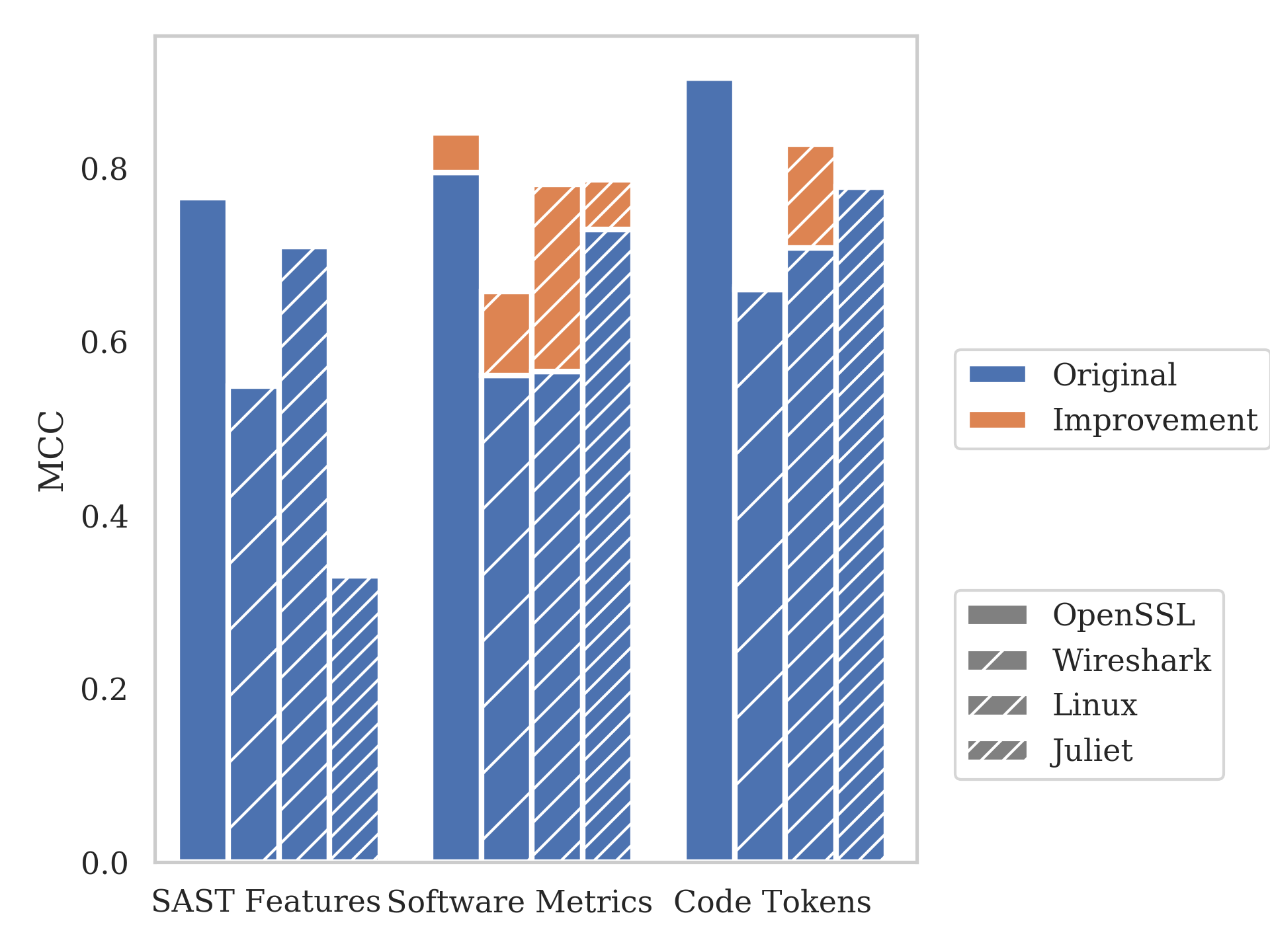}
  \caption{Model performance improvement for the addition of SAST features.}
  \label{fig:sast_impact}
  \Description{A bar chart with MCC improvement on the y-axis and feature types on the x-axis for the four datasets.}
\end{figure}

For our two main feature sets (software metrics and code tokens), we inspected model performance increase of adding SAST based features to SVP models, shown in Figure \ref{fig:sast_impact}. We also considered the use of SAST features alone (the left-hand values of Figure \ref{fig:sast_impact}) to represent the scenario in which we  used learning-based approaches to enhance the SAST tool outputs, through the reduction of false positives \cite{pereira2019, yoon2014}. Like the previous studies on such topics, we also observed that this approach can improve the performance of SAST tools. However, SAST features alone did not outperform regular SVP models, despite introducing the same weakness of data requirements and coarse granularity. These approaches only serve to help reduce the gap in performance between SAST and SVP. They do not reduce false negatives, hence, they are unable to produce a approach that is greater than the sum of its parts. 

A more promising approach is to use SAST features in combination with regular SVP models. As seen in Figure \ref{fig:sast_impact}, SAST features were able to improve the software metrics model due to their similar individual performances. However, SAST features still performed worse than the code tokens (with the exception of the Linux dataset). 

Whilst the unification of two approaches (i.e., SAST tools and SVPs) appeared useful in some circumstances, it was generally unable to elevate the potential of SVP. For the use of a combined approach to be cost effective for an organisation, we would require their combined use to contribute more value than using one of the approaches individually, which we find not to be the case in terms of detection performance. Hence, our investigation of the potential viability of unifying the two approaches concluded insufficient outcomes; that means there is a need of developing more sophisticated approaches in the future. 

\begin{tcolorbox}[right=1pt, left=1pt, top=1pt, bottom=1pt, colback=white]
    \textbf{Finding 12:} \textit{The two approaches lack synergy. It is difficult to improve the overall performance through unification.}
\end{tcolorbox}

\section{Discussion}
This study aimed at empirically investigating the capabilities of learning-based approaches for SAST in comparison to rule-based tools for determining the potential synergies between them. Through RQ1 and RQ2, we have found that SVP models exhibit better \textit{overall} performance for both detection and assessment. However, as SVP models have some caveats that SAST tools do not, such as stringent data requirements, coarse granularity, and poor transparency, SAST tools can still be used to achieve recall rates similar to the learning-based approaches. For RQ3 however, we have identified that these two approaches actually lack synergy, through the insignificant and even negative results we obtained for their unification. 

However, both of the compared SAST approaches are far from achieving specific SV detection and localization. The actual vulnerable lines and components of a file are usually small (1-5 lines), but the average open source code file is very large (1000+ lines). Hence, inspection from a file-level warning requires significant effort. Whilst SAST tools produce raw warnings at the line-level, the warnings are not accurate; there is a trade-off between granularity and effectiveness. Without effective localization, SAST approaches are limited in their actual testing capabilities. They are best used to assist with manual inspection or to direct more expensive dynamic testing efforts. Whilst this is useful for reducing the effort required for software quality assurance, developers still desire stronger capabilities from source code security analysis approaches in the modern agile software development paradigms \cite{oyetoyan2018}. 

\subsection{Observations on Method Usability}
Developers also require tools to be convenient. Whilst tool performance is the most important capability of a SAST approach, the usability of tools also serve as barriers to adoption \cite{christakis2016,johnson2013,oyetoyan2018}. Hence, to provide an additional view of each approaches' capabilities for adoption, we contribute a discussion of their comparative convenience and ease of application gained through our experiences from conducting this empirical study. We frame our discussion based on the desired properties outlined by Poel \cite{poel2010}. 

\textbf{\textit{Responsiveness.}} The responsiveness of an approach is the required run-time to scan files. The pattern matching based SAST tools (Flawfinder and RATS) only took a few minutes to scan the datasets. This speed does not slow down or require stopping of development processes, which is optimal \cite{christakis2016}. SVP models operated even faster, taking an average of 6 seconds to make their predictions. However, this required the model to already be setup, which is a time consuming process (training time took an average of 86 minutes per model). Cppcheck, which uses a more sophisticated data flow analysis method, was considerably slower as it took an average of a few hours to scan the entire dataset, which is not scalable. Codebase scans would need to be conducted overnight or at implementation stopping points, making integration difficult.  

\textbf{\textit{Error-proneness.}} During our benchmarking process, Flawfinder produced several run-time errors, but was able to complete its scan. However, Cppcheck occasionally crashed due to code scanning errors. This error-proneness is likely from the generalisability of these tools, which are designed to operate on heterogeneous codebases. Hence, there will inevitably be errors for certain applications' contexts. SVP models did not produce any errors during run-time as they are built towards their respective context. 

\textbf{\textit{Interpretability.}} Interpretability has two dimensions \cite{mi2020}: the local interpretability of individual predictions, and the global interpretability of the overall tool/model. In terms of local interpretability, SAST tools provide a description for each flagged warning. This makes the warnings more actionable by developers during manual verification, as it assists with understanding and mitigation. SAST tools also have global interpretability as their approaches are transparent. Their documentation describe the rules used to produce warnings, and the checks they are capable of. The SVP models we implemented were not interpretable or transparent (\textit{Finding 7}). Although methods exist for interpretable machine learning, we did not incorporate these as they have not yet been properly explored in existing SVP literature \cite{zeng2020}. Making SVP models interpretable is important however, to increase model trust and actionability \cite{tantithamthavorn2020}. 

\textbf{\textit{Setup.}} For SAST tools, we have found that the setup and configuration were generally easy. All three tools had minor customizability. However, this may only be reflected in open source tools, as the prior studies have found the configuration to be a pain point \cite{christakis2016,johnson2013,oyetoyan2018}. We found the setup and configuration to be much more of a challenge for SVP models. Although the implementation itself is relatively straightforward, the construction also needs model requirement identification, tuning, and validation, which would put further strain on developers. The optimisation requirements of SVP model setup are particularly important, as we observed the prediction performance to vary heavily across different classifier algorithms and whilst tuning. 

\subsection{Implications}
Our findings have produced implications for both developers and researchers, based on the discovered limitations of each approach. 

\subsubsection{Developers}
For developers, we provide insights into the comparative performance and use of the investigated SAST approaches. Additionally, we propose some preliminary recommendations for the use of these approaches  based on our findings. 

We suggest that these two approaches for source code security analysis should be used separately, at least until better approaches of unification are developed. Tool integration is expensive \cite{christakis2016,johnson2013}, and hence most organisations will desire to only adopt a singular approach. From \textit{Finding 3 \& 12}, we find that some of the detection capabilities overlap and hence one approach is sufficient to replace the other. As we have identified in \textit{Finding 8}, SAST tools are a good initial option for timely software quality assurance, but SVP models have shown better performance as a project matures. 

If SAST tools are used, they should be applied to a reduced range of code, rather than to all files iteratively. In \textit{Finding 2 \& 5}, we have identified that SAST tools incur heavy inspection costs due to their high false positive rates. Furthermore, alongside \textit{Finding 6}, we find that SAST tools struggle to discern code or security context themselves. Hence, to reduce the subsequent inspection efforts, developers should only use SAST tools on areas of code that they are willing to manually inspect or test, and should not apply these tools to code they confidently consider secure. 

SVP models should be regularly retrained and updated. From \textit{Finding 8}, we observe that SVP models can be adopted relatively early. On the OpenSSL dataset, we found that even when using just 110 files for training (35 labelled as vulnerable), the model still achieved \textgreater0.7 MCC. However, we note that SVP model performance increasingly improves as more data is used to train the model, and SV assessment is also constrained to the training data (\textit{Finding 11}); hence, models should be regularly updated. 

\subsubsection{Researchers}
For researchers, we have identified some promising research directions based on the several pain points we have identified for each of the compared approaches. 

For SAST tools, the main pain point is their performance. Whilst their capabilities for detection are decent (\textit{Finding 1}), they produce a large number of false positives (\textit{Finding 2}). Naturally, researchers and tool-makers should continue to develop better performing methods. We observe that SAST tools exhibit a trade-off between precision and recall (\textit{Finding 4}), so overcoming this hurdle is a must. 

For SVP models, the main pain points come from the required knowledge and experience of using them, rather than performance. SVP models require more transparency (\textit{Finding 7}); developers need to be able to understand the capabilities and limitations of the software quality assurance approaches they use. Additionally, local prediction interpretation will assist developers in the inspection of the predicted modules. Many approaches for interpretable machine learning exist \cite{mi2020}, but they have so far been under-explored in the context of SVP. More research needs to be conducted in this area. 

Additionally, SVP models need better approaches for localization. Although SVP models produce a lower number of files to inspect than SAST tools (\textit{Finding 5}), they still have high inspection costs due to the lack of localization. Models have been developed that predict vulnerable components at finer levels of granularity, through code slices \cite{li2018} or commits \cite{perl2015}. However, these techniques often require the enclosing context or scope of a vulnerability \cite{sahal2018}, and hence are not perfect solutions. Recent techniques have been proposed for fault localization of defect prediction models \cite{pornprasit2021}. Researchers should continue to advance these approaches and identify how the approaches transfer to SVP. 

Finally, more focus needs to be put on the assessment task. SV assessment is an important task for SV mitigation, but it is not considered as a main priority for either of the approaches, hence, the performance suffers (\textit{Finding 9}). More focus needs to be put into this task to achieve the performance that is more consistent with that of SV detection. Another important component of SV assessment is prioritization; assessing the risk of each SV. Although SAST tools often produce an indicative ranking of their warnings, these are not based on the actual severity or exploitability of SVs. For SVP models, little work has been conducted on using learning-based approaches to predict SV risk at the source code level. 

\section{Threats to Validity}
\textit{External Validity.} We only investigated 3 open source projects and SARD test cases for vulnerabilities of the C/C++ programming language. We acknowledge that our findings may not generalize to other projects or programming languages. 

\textit{Internal Validity.} Our tuning of SVP models and configuration of SAST tools are potentially sub-optimal. To lessen this threat, we tuned a wide range of hyperparameters for our SVP models to optimize them in relation to our dataset. For the SAST tools, we manually analyzed the configuration options and initial outputs to determine the best configuration. 

\textit{Construct Validity.} The range of tools and models we considered is imperfect. We only selected open source SAST tools, as these are the most readily available and widely used. Commercial tools which use more sophisticated techniques will likely produce different performances \cite{aloraini2019}. Concurrently, the three SVP models we built were also relatively basic. However, selecting simplistic base approaches gives us a more general view for comparison, as they are the most common representation. 
Our open source datasets only contained documented post-release vulnerabilities from NVD. This largely conceals vulnerabilities already detected or removed during the implementation phase from our analysis. Similarly, we have found that all three of our open source projects have existing documentation of SAST tool usage, including \textit{Cppcheck}, but the usage of these tools is not consistent or thorough\footnote{\url{https://github.com/openssl/openssl/issues/5013}}. To help overcome this limitation, we also evaluated the compared approaches on the Juliet Test Suite \cite{boland2012}, to obtain a more complete view of development vulnerabilities.
Our evaluation of SAST tools and SVP approaches included qualitative assessment and manual analysis. Such investigation has the potential of being impacted by subjectivity and human bias. To ensure more reliable human evaluation, we used multiple assessors in this study. 

\textit{Conclusion Validity.} To help strengthen conclusion validity, we confirmed our results using non-parametric statistical tests \cite{wilcoxon1992}, and did inspection on statistically significant sample sizes \cite{cochran2007}. 

\section{Conclusion}
This study has conducted the first large-scale comparative analysis of rule-based (SAST tools) and learning-based (SVP models) approaches to source code security analysis. Through this analysis, we have identified their comparative capabilities and uses, which we present through 12 main findings. From the findings of this study, we have also derived several implications for both researchers and practitioners to support the selection and use of SAST approaches, as well as direct future research efforts in this area. We conclude that SAST tools and SVP models provide similar detection capabilities, but SVP models provide better overall performance for SV detection and assessment. However, SVP models exhibit some caveats that SAST tools do not, such as data requirements, coarse granularity, and difficult interpretation. 

In the future, we aim to conduct a user-survey to see how developers consider the comparative trade-offs that we have identified. 

\begin{acks}
This work has been supported by the Cyber Security Cooperative Research Centre Limited whose activities are partially funded by the Australian Government’s Cooperative Research Centre Programme. 
\end{acks}

\bibliographystyle{ACM-Reference-Format}
\bibliography{bibfile}


\begin{thebibliography}{68}


\ifx \showCODEN    \undefined \def \showCODEN     #1{\unskip}     \fi
\ifx \showDOI      \undefined \def \showDOI       #1{#1}\fi
\ifx \showISBNx    \undefined \def \showISBNx     #1{\unskip}     \fi
\ifx \showISBNxiii \undefined \def \showISBNxiii  #1{\unskip}     \fi
\ifx \showISSN     \undefined \def \showISSN      #1{\unskip}     \fi
\ifx \showLCCN     \undefined \def \showLCCN      #1{\unskip}     \fi
\ifx \shownote     \undefined \def \shownote      #1{#1}          \fi
\ifx \showarticletitle \undefined \def \showarticletitle #1{#1}   \fi
\ifx \showURL      \undefined \def \showURL       {\relax}        \fi
\providecommand\bibfield[2]{#2}
\providecommand\bibinfo[2]{#2}
\providecommand\natexlab[1]{#1}
\providecommand\showeprint[2][]{arXiv:#2}

\bibitem[\protect\citeauthoryear{Aloraini, Nagappan, German, Hayashi, and
  Higo}{Aloraini et~al\mbox{.}}{2019}]%
        {aloraini2019}
\bibfield{author}{\bibinfo{person}{Bushra Aloraini}, \bibinfo{person}{Meiyappan
  Nagappan}, \bibinfo{person}{Daniel~M German}, \bibinfo{person}{Shinpei
  Hayashi}, {and} \bibinfo{person}{Yoshiki Higo}.}
  \bibinfo{year}{2019}\natexlab{}.
\newblock \showarticletitle{An empirical study of security warnings from static
  application security testing tools}.
\newblock \bibinfo{journal}{\emph{Journal of Systems and Software}}
  \bibinfo{volume}{158} (\bibinfo{year}{2019}), \bibinfo{pages}{110427}.
\newblock


\bibitem[\protect\citeauthoryear{Beller, Bholanath, McIntosh, and
  Zaidman}{Beller et~al\mbox{.}}{2016}]%
        {beller2016}
\bibfield{author}{\bibinfo{person}{Moritz Beller}, \bibinfo{person}{Radjino
  Bholanath}, \bibinfo{person}{Shane McIntosh}, {and} \bibinfo{person}{Andy
  Zaidman}.} \bibinfo{year}{2016}\natexlab{}.
\newblock \showarticletitle{Analyzing the state of static analysis: A
  large-scale evaluation in open source software}. In
  \bibinfo{booktitle}{\emph{2016 IEEE 23rd International Conference on Software
  Analysis, Evolution, and Reengineering (SANER)}}, Vol.~\bibinfo{volume}{1}.
  IEEE, \bibinfo{pages}{470--481}.
\newblock


\bibitem[\protect\citeauthoryear{Boland and Black}{Boland and Black}{2012}]%
        {boland2012}
\bibfield{author}{\bibinfo{person}{Tim Boland} {and} \bibinfo{person}{Paul~E
  Black}.} \bibinfo{year}{2012}\natexlab{}.
\newblock \showarticletitle{Juliet 1.1 C/C++ and Java test suite}.
\newblock \bibinfo{journal}{\emph{IEEE Computer Architecture Letters}}
  \bibinfo{volume}{45}, \bibinfo{number}{10} (\bibinfo{year}{2012}),
  \bibinfo{pages}{88--90}.
\newblock


\bibitem[\protect\citeauthoryear{CERN}{CERN}{[n.d.]}]%
        {rats}
\bibfield{author}{\bibinfo{person}{CERN}.} \bibinfo{year}{[n.d.]}\natexlab{}.
\newblock \bibinfo{title}{Rough Auditing Tool for Security (RATS)}.
\newblock
\newblock
\urldef\tempurl%
\url{https://security.web.cern.ch/recommendations/en/codetools/rats.shtml}
\showURL{%
\tempurl}


\bibitem[\protect\citeauthoryear{Chicco and Jurman}{Chicco and Jurman}{2020}]%
        {chicco2020}
\bibfield{author}{\bibinfo{person}{Davide Chicco} {and}
  \bibinfo{person}{Giuseppe Jurman}.} \bibinfo{year}{2020}\natexlab{}.
\newblock \showarticletitle{The advantages of the Matthews correlation
  coefficient (MCC) over F1 score and accuracy in binary classification
  evaluation}.
\newblock \bibinfo{journal}{\emph{BMC genomics}} \bibinfo{volume}{21},
  \bibinfo{number}{1} (\bibinfo{year}{2020}), \bibinfo{pages}{1--13}.
\newblock


\bibitem[\protect\citeauthoryear{Chowdhury and Zulkernine}{Chowdhury and
  Zulkernine}{2011}]%
        {chowdhury2011}
\bibfield{author}{\bibinfo{person}{Istehad Chowdhury} {and}
  \bibinfo{person}{Mohammad Zulkernine}.} \bibinfo{year}{2011}\natexlab{}.
\newblock \showarticletitle{Using complexity, coupling, and cohesion metrics as
  early indicators of vulnerabilities}.
\newblock \bibinfo{journal}{\emph{Journal of Systems Architecture}}
  \bibinfo{volume}{57}, \bibinfo{number}{3} (\bibinfo{year}{2011}),
  \bibinfo{pages}{294--313}.
\newblock


\bibitem[\protect\citeauthoryear{Christakis and Bird}{Christakis and
  Bird}{2016}]%
        {christakis2016}
\bibfield{author}{\bibinfo{person}{Maria Christakis} {and}
  \bibinfo{person}{Christian Bird}.} \bibinfo{year}{2016}\natexlab{}.
\newblock \showarticletitle{What developers want and need from program
  analysis: an empirical study}. In \bibinfo{booktitle}{\emph{Proceedings of
  the 31st IEEE/ACM international conference on automated software
  engineering}}. \bibinfo{pages}{332--343}.
\newblock


\bibitem[\protect\citeauthoryear{Cochran}{Cochran}{2007}]%
        {cochran2007}
\bibfield{author}{\bibinfo{person}{William~G Cochran}.}
  \bibinfo{year}{2007}\natexlab{}.
\newblock \bibinfo{booktitle}{\emph{Sampling techniques}}.
\newblock \bibinfo{publisher}{John Wiley \& Sons}.
\newblock


\bibitem[\protect\citeauthoryear{Coulter, Han, Pan, Zhang, and Xiang}{Coulter
  et~al\mbox{.}}{2020}]%
        {coulter2020}
\bibfield{author}{\bibinfo{person}{Rory Coulter}, \bibinfo{person}{Qing-Long
  Han}, \bibinfo{person}{Lei Pan}, \bibinfo{person}{Jun Zhang}, {and}
  \bibinfo{person}{Yang Xiang}.} \bibinfo{year}{2020}\natexlab{}.
\newblock \showarticletitle{Code analysis for intelligent cyber systems: A
  data-driven approach}.
\newblock \bibinfo{journal}{\emph{Information sciences}}  \bibinfo{volume}{524}
  (\bibinfo{year}{2020}), \bibinfo{pages}{46--58}.
\newblock


\bibitem[\protect\citeauthoryear{Croft, Newlands, Chen, and Babar}{Croft
  et~al\mbox{.}}{2021}]%
        {reproduction_package}
\bibfield{author}{\bibinfo{person}{Roland Croft}, \bibinfo{person}{Dominic
  Newlands}, \bibinfo{person}{Ziyu Chen}, {and} \bibinfo{person}{Ali Babar}.}
  \bibinfo{year}{2021}\natexlab{}.
\newblock \bibinfo{title}{Reproduction package for "An Empirical Study of
  Rule-Based and Learning-Based Approaches for Static Application Security
  Testing"}.
\newblock
\newblock
\urldef\tempurl%
\url{https://doi.org/10.6084/m9.figshare.14585076.v1}
\showDOI{\tempurl}


\bibitem[\protect\citeauthoryear{CWE}{CWE}{[n.d.]}]%
        {CWE}
\bibfield{author}{\bibinfo{person}{CWE}.} \bibinfo{year}{[n.d.]}\natexlab{}.
\newblock \bibinfo{title}{Common Weakness Enumeration}.
\newblock
\newblock
\urldef\tempurl%
\url{https://cwe.mitre.org/}
\showURL{%
\tempurl}


\bibitem[\protect\citeauthoryear{D{\'\i}az and Bermejo}{D{\'\i}az and
  Bermejo}{2013}]%
        {diaz2013}
\bibfield{author}{\bibinfo{person}{Gabriel D{\'\i}az} {and}
  \bibinfo{person}{Juan~Ram{\'o}n Bermejo}.} \bibinfo{year}{2013}\natexlab{}.
\newblock \showarticletitle{Static analysis of source code security: Assessment
  of tools against SAMATE tests}.
\newblock \bibinfo{journal}{\emph{Information and software technology}}
  \bibinfo{volume}{55}, \bibinfo{number}{8} (\bibinfo{year}{2013}),
  \bibinfo{pages}{1462--1476}.
\newblock


\bibitem[\protect\citeauthoryear{Falessi, Huang, Narayana, Thai, and
  Turhan}{Falessi et~al\mbox{.}}{2020}]%
        {falessi2020}
\bibfield{author}{\bibinfo{person}{Davide Falessi}, \bibinfo{person}{Jacky
  Huang}, \bibinfo{person}{Likhita Narayana}, \bibinfo{person}{Jennifer~Fong
  Thai}, {and} \bibinfo{person}{Burak Turhan}.}
  \bibinfo{year}{2020}\natexlab{}.
\newblock \showarticletitle{On the need of preserving order of data when
  validating within-project defect classifiers}.
\newblock \bibinfo{journal}{\emph{Empirical Software Engineering}}
  \bibinfo{volume}{25}, \bibinfo{number}{6} (\bibinfo{year}{2020}),
  \bibinfo{pages}{4805--4830}.
\newblock


\bibitem[\protect\citeauthoryear{Fan, da~Costa, Lo, Hassan, and Shanping}{Fan
  et~al\mbox{.}}{2020}]%
        {fan2020}
\bibfield{author}{\bibinfo{person}{Yuanrui Fan}, \bibinfo{person}{D~Alencar da
  Costa}, \bibinfo{person}{D Lo}, \bibinfo{person}{AE Hassan}, {and}
  \bibinfo{person}{L Shanping}.} \bibinfo{year}{2020}\natexlab{}.
\newblock \showarticletitle{The impact of mislabeled changes by szz on
  just-in-time defect prediction}.
\newblock \bibinfo{journal}{\emph{IEEE Transactions on Software Engineering}}
  (\bibinfo{year}{2020}).
\newblock


\bibitem[\protect\citeauthoryear{Foundation}{Foundation}{[n.d.]}]%
        {OWASP_static}
\bibfield{author}{\bibinfo{person}{OWASP Foundation}.}
  \bibinfo{year}{[n.d.]}\natexlab{}.
\newblock \bibinfo{title}{Static Code Analysis}.
\newblock
\newblock
\urldef\tempurl%
\url{https://owasp.org/www-community/controls/Static_Code_Analysis}
\showURL{%
\tempurl}


\bibitem[\protect\citeauthoryear{Gegick and Williams}{Gegick and
  Williams}{2007}]%
        {gegick2007}
\bibfield{author}{\bibinfo{person}{Michael Gegick} {and}
  \bibinfo{person}{Laurie Williams}.} \bibinfo{year}{2007}\natexlab{}.
\newblock \showarticletitle{Toward the use of automated static analysis alerts
  for early identification of vulnerability-and attack-prone components}. In
  \bibinfo{booktitle}{\emph{Second International Conference on Internet
  Monitoring and Protection (ICIMP 2007)}}. IEEE, \bibinfo{pages}{18--18}.
\newblock


\bibitem[\protect\citeauthoryear{Ghaffarian and Shahriari}{Ghaffarian and
  Shahriari}{2017}]%
        {ghaffarian2017}
\bibfield{author}{\bibinfo{person}{Seyed~Mohammad Ghaffarian} {and}
  \bibinfo{person}{Hamid~Reza Shahriari}.} \bibinfo{year}{2017}\natexlab{}.
\newblock \showarticletitle{Software vulnerability analysis and discovery using
  machine-learning and data-mining techniques: A survey}.
\newblock \bibinfo{journal}{\emph{ACM Computing Surveys (CSUR)}}
  \bibinfo{volume}{50}, \bibinfo{number}{4} (\bibinfo{year}{2017}),
  \bibinfo{pages}{1--36}.
\newblock


\bibitem[\protect\citeauthoryear{Ghotra, McIntosh, and Hassan}{Ghotra
  et~al\mbox{.}}{2017}]%
        {ghotra2017}
\bibfield{author}{\bibinfo{person}{Baljinder Ghotra}, \bibinfo{person}{Shane
  McIntosh}, {and} \bibinfo{person}{Ahmed~E Hassan}.}
  \bibinfo{year}{2017}\natexlab{}.
\newblock \showarticletitle{A large-scale study of the impact of feature
  selection techniques on defect classification models}. In
  \bibinfo{booktitle}{\emph{2017 IEEE/ACM 14th International Conference on
  Mining Software Repositories (MSR)}}. IEEE, \bibinfo{pages}{146--157}.
\newblock


\bibitem[\protect\citeauthoryear{Hanif, Nasir, Ab~Razak, Firdaus, and
  Anuar}{Hanif et~al\mbox{.}}{2021}]%
        {hanif2021}
\bibfield{author}{\bibinfo{person}{Hazim Hanif}, \bibinfo{person}{Mohd Hairul
  Nizam~Md Nasir}, \bibinfo{person}{Mohd~Faizal Ab~Razak},
  \bibinfo{person}{Ahmad Firdaus}, {and} \bibinfo{person}{Nor~Badrul Anuar}.}
  \bibinfo{year}{2021}\natexlab{}.
\newblock \showarticletitle{The rise of software vulnerability: Taxonomy of
  software vulnerabilities detection and machine learning approaches}.
\newblock \bibinfo{journal}{\emph{Journal of Network and Computer
  Applications}} (\bibinfo{year}{2021}), \bibinfo{pages}{103009}.
\newblock


\bibitem[\protect\citeauthoryear{Imtiaz, Rahman, Farhana, and Williams}{Imtiaz
  et~al\mbox{.}}{2019}]%
        {imtiaz2019}
\bibfield{author}{\bibinfo{person}{Nasif Imtiaz}, \bibinfo{person}{Akond
  Rahman}, \bibinfo{person}{Effat Farhana}, {and} \bibinfo{person}{Laurie
  Williams}.} \bibinfo{year}{2019}\natexlab{}.
\newblock \showarticletitle{Challenges with responding to static analysis tool
  alerts}. In \bibinfo{booktitle}{\emph{2019 IEEE/ACM 16th International
  Conference on Mining Software Repositories (MSR)}}. IEEE,
  \bibinfo{pages}{245--249}.
\newblock


\bibitem[\protect\citeauthoryear{Jaccard}{Jaccard}{1912}]%
        {jaccard1912}
\bibfield{author}{\bibinfo{person}{Paul Jaccard}.}
  \bibinfo{year}{1912}\natexlab{}.
\newblock \showarticletitle{The distribution of the flora in the alpine zone.
  1}.
\newblock \bibinfo{journal}{\emph{New phytologist}} \bibinfo{volume}{11},
  \bibinfo{number}{2} (\bibinfo{year}{1912}), \bibinfo{pages}{37--50}.
\newblock


\bibitem[\protect\citeauthoryear{Jimenez, Le~Traon, and Papadakis}{Jimenez
  et~al\mbox{.}}{2018}]%
        {jimenez2018}
\bibfield{author}{\bibinfo{person}{Matthieu Jimenez}, \bibinfo{person}{Yves
  Le~Traon}, {and} \bibinfo{person}{Mike Papadakis}.}
  \bibinfo{year}{2018}\natexlab{}.
\newblock \showarticletitle{Enabling the continous analysis of security
  vulnerabilities with vuldata7}. In \bibinfo{booktitle}{\emph{IEEE
  International Working Conference on Source Code Analysis and Manipulation}}.
\newblock


\bibitem[\protect\citeauthoryear{Johnson, Song, Murphy-Hill, and
  Bowdidge}{Johnson et~al\mbox{.}}{2013}]%
        {johnson2013}
\bibfield{author}{\bibinfo{person}{Brittany Johnson}, \bibinfo{person}{Yoonki
  Song}, \bibinfo{person}{Emerson Murphy-Hill}, {and} \bibinfo{person}{Robert
  Bowdidge}.} \bibinfo{year}{2013}\natexlab{}.
\newblock \showarticletitle{Why don't software developers use static analysis
  tools to find bugs?}. In \bibinfo{booktitle}{\emph{2013 35th International
  Conference on Software Engineering (ICSE)}}. IEEE, \bibinfo{pages}{672--681}.
\newblock


\bibitem[\protect\citeauthoryear{Kaur and Nayyar}{Kaur and Nayyar}{2020}]%
        {kaur2020}
\bibfield{author}{\bibinfo{person}{Arvinder Kaur} {and}
  \bibinfo{person}{Ruchikaa Nayyar}.} \bibinfo{year}{2020}\natexlab{}.
\newblock \showarticletitle{A comparative study of static code analysis tools
  for vulnerability detection in c/c++ and java source code}.
\newblock \bibinfo{journal}{\emph{Procedia Computer Science}}
  \bibinfo{volume}{171} (\bibinfo{year}{2020}), \bibinfo{pages}{2023--2029}.
\newblock


\bibitem[\protect\citeauthoryear{Kendall}{Kendall}{1938}]%
        {kendall1938}
\bibfield{author}{\bibinfo{person}{Maurice~G Kendall}.}
  \bibinfo{year}{1938}\natexlab{}.
\newblock \showarticletitle{A new measure of rank correlation}.
\newblock \bibinfo{journal}{\emph{Biometrika}} \bibinfo{volume}{30},
  \bibinfo{number}{1/2} (\bibinfo{year}{1938}), \bibinfo{pages}{81--93}.
\newblock


\bibitem[\protect\citeauthoryear{Khan and Parkinson}{Khan and
  Parkinson}{2018}]%
        {khan2018}
\bibfield{author}{\bibinfo{person}{Saad Khan} {and} \bibinfo{person}{Simon
  Parkinson}.} \bibinfo{year}{2018}\natexlab{}.
\newblock \showarticletitle{Review into state of the art of vulnerability
  assessment using artificial intelligence}.
\newblock In \bibinfo{booktitle}{\emph{Guide to Vulnerability Analysis for
  Computer Networks and Systems}}. \bibinfo{publisher}{Springer},
  \bibinfo{pages}{3--32}.
\newblock


\bibitem[\protect\citeauthoryear{Kildall}{Kildall}{1973}]%
        {kildall1973}
\bibfield{author}{\bibinfo{person}{Gary~A Kildall}.}
  \bibinfo{year}{1973}\natexlab{}.
\newblock \showarticletitle{A unified approach to global program optimization}.
  In \bibinfo{booktitle}{\emph{Proceedings of the 1st annual ACM SIGACT-SIGPLAN
  symposium on Principles of programming languages}}.
  \bibinfo{pages}{194--206}.
\newblock


\bibitem[\protect\citeauthoryear{Le, Hin, Croft, and Babar}{Le
  et~al\mbox{.}}{2020}]%
        {le2020}
\bibfield{author}{\bibinfo{person}{Triet Huynh~Minh Le}, \bibinfo{person}{David
  Hin}, \bibinfo{person}{Roland Croft}, {and} \bibinfo{person}{M~Ali Babar}.}
  \bibinfo{year}{2020}\natexlab{}.
\newblock \showarticletitle{PUMiner: Mining Security Posts from Developer
  Question and Answer Websites with PU Learning}. In
  \bibinfo{booktitle}{\emph{Proceedings of the 17th International Conference on
  Mining Software Repositories}}. \bibinfo{pages}{350--361}.
\newblock


\bibitem[\protect\citeauthoryear{Le~Huynh~Minh, Croft, Hin, and
  Ali~Babar}{Le~Huynh~Minh et~al\mbox{.}}{2021}]%
        {le2021}
\bibfield{author}{\bibinfo{person}{Triet~Le Le~Huynh~Minh},
  \bibinfo{person}{Roland Croft}, \bibinfo{person}{David Hin}, {and}
  \bibinfo{person}{Muhammad~Ali Ali~Babar}.} \bibinfo{year}{2021}\natexlab{}.
\newblock \showarticletitle{A Large-scale Study of Security Vulnerability
  Support on Developer Q\&A Websites}.
\newblock In \bibinfo{booktitle}{\emph{Evaluation and Assessment in Software
  Engineering}}. \bibinfo{pages}{109--118}.
\newblock


\bibitem[\protect\citeauthoryear{Li, Zou, Xu, Ou, Jin, Wang, Deng, and
  Zhong}{Li et~al\mbox{.}}{2018}]%
        {li2018}
\bibfield{author}{\bibinfo{person}{Zhen Li}, \bibinfo{person}{Deqing Zou},
  \bibinfo{person}{Shouhuai Xu}, \bibinfo{person}{Xinyu Ou},
  \bibinfo{person}{Hai Jin}, \bibinfo{person}{Sujuan Wang},
  \bibinfo{person}{Zhijun Deng}, {and} \bibinfo{person}{Yuyi Zhong}.}
  \bibinfo{year}{2018}\natexlab{}.
\newblock \showarticletitle{Vuldeepecker: A deep learning-based system for
  vulnerability detection}. In \bibinfo{booktitle}{\emph{25th Annual Network
  and Distributed System Symposium}}.
\newblock


\bibitem[\protect\citeauthoryear{Marjamaki}{Marjamaki}{[n.d.]}]%
        {cppcheck}
\bibfield{author}{\bibinfo{person}{Daniel Marjamaki}.}
  \bibinfo{year}{[n.d.]}\natexlab{}.
\newblock \bibinfo{title}{Cppcheck}.
\newblock
\newblock
\urldef\tempurl%
\url{http://cppcheck.sourceforge.net/}
\showURL{%
\tempurl}


\bibitem[\protect\citeauthoryear{Mi, Li, and Zhou}{Mi et~al\mbox{.}}{2020}]%
        {mi2020}
\bibfield{author}{\bibinfo{person}{Jian-Xun Mi}, \bibinfo{person}{An-Di Li},
  {and} \bibinfo{person}{Li-Fang Zhou}.} \bibinfo{year}{2020}\natexlab{}.
\newblock \showarticletitle{Review Study of Interpretation Methods for Future
  Interpretable Machine Learning}.
\newblock \bibinfo{journal}{\emph{IEEE Access}}  \bibinfo{volume}{8}
  (\bibinfo{year}{2020}), \bibinfo{pages}{191969--191985}.
\newblock


\bibitem[\protect\citeauthoryear{Morrison, Herzig, Murphy, and
  Williams}{Morrison et~al\mbox{.}}{2015}]%
        {morrison2015}
\bibfield{author}{\bibinfo{person}{Patrick Morrison}, \bibinfo{person}{Kim
  Herzig}, \bibinfo{person}{Brendan Murphy}, {and} \bibinfo{person}{Laurie
  Williams}.} \bibinfo{year}{2015}\natexlab{}.
\newblock \showarticletitle{Challenges with applying vulnerability prediction
  models}. In \bibinfo{booktitle}{\emph{Proceedings of the 2015 Symposium and
  Bootcamp on the Science of Security}}. \bibinfo{pages}{1--9}.
\newblock


\bibitem[\protect\citeauthoryear{Morrison, Pandita, Xiao, Chillarege, and
  Williams}{Morrison et~al\mbox{.}}{2018}]%
        {morrison2018}
\bibfield{author}{\bibinfo{person}{Patrick~J Morrison}, \bibinfo{person}{Rahul
  Pandita}, \bibinfo{person}{Xusheng Xiao}, \bibinfo{person}{Ram Chillarege},
  {and} \bibinfo{person}{Laurie Williams}.} \bibinfo{year}{2018}\natexlab{}.
\newblock \showarticletitle{Are vulnerabilities discovered and resolved like
  other defects?}
\newblock \bibinfo{journal}{\emph{Empirical Software Engineering}}
  \bibinfo{volume}{23}, \bibinfo{number}{3} (\bibinfo{year}{2018}),
  \bibinfo{pages}{1383--1421}.
\newblock


\bibitem[\protect\citeauthoryear{Munaiah and Meneely}{Munaiah and
  Meneely}{2019}]%
        {munaiah2019}
\bibfield{author}{\bibinfo{person}{Nuthan Munaiah} {and}
  \bibinfo{person}{Andrew Meneely}.} \bibinfo{year}{2019}\natexlab{}.
\newblock \showarticletitle{Data-driven insights from vulnerability discovery
  metrics}. In \bibinfo{booktitle}{\emph{2019 IEEE/ACM Joint 4th International
  Workshop on Rapid Continuous Software Engineering and 1st International
  Workshop on Data-Driven Decisions, Experimentation and Evolution
  (RCoSE/DDrEE)}}. IEEE, \bibinfo{pages}{1--7}.
\newblock


\bibitem[\protect\citeauthoryear{Mushtaq, Rasool, and Shehzad}{Mushtaq
  et~al\mbox{.}}{2017}]%
        {mushtaq2017}
\bibfield{author}{\bibinfo{person}{Zaigham Mushtaq}, \bibinfo{person}{Ghulam
  Rasool}, {and} \bibinfo{person}{Balawal Shehzad}.}
  \bibinfo{year}{2017}\natexlab{}.
\newblock \showarticletitle{Multilingual source code analysis: A systematic
  literature review}.
\newblock \bibinfo{journal}{\emph{IEEE Access}}  \bibinfo{volume}{5}
  (\bibinfo{year}{2017}), \bibinfo{pages}{11307--11336}.
\newblock


\bibitem[\protect\citeauthoryear{of~Standards and Technology}{of~Standards and
  Technology}{[n.d.]a}]%
        {SARD}
\bibfield{author}{\bibinfo{person}{National~Institute of Standards} {and}
  \bibinfo{person}{Technology}.} \bibinfo{year}{[n.d.]}\natexlab{a}.
\newblock \bibinfo{title}{Software Assurance and Reference Dataset}.
\newblock
\newblock
\urldef\tempurl%
\url{https://samate.nist.gov/SARD/testsuite.php}
\showURL{%
\tempurl}


\bibitem[\protect\citeauthoryear{of~Standards and Technology}{of~Standards and
  Technology}{[n.d.]b}]%
        {NIST_tools}
\bibfield{author}{\bibinfo{person}{National~Institute of Standards} {and}
  \bibinfo{person}{Technology}.} \bibinfo{year}{[n.d.]}\natexlab{b}.
\newblock \bibinfo{title}{Source Code Security Analyzers}.
\newblock
\newblock
\urldef\tempurl%
\url{https://samate.nist.gov/index.php/Source_Code_Security_Analyzers.html}
\showURL{%
\tempurl}


\bibitem[\protect\citeauthoryear{Oyetoyan, Milosheska, Grini, and
  Cruzes}{Oyetoyan et~al\mbox{.}}{2018}]%
        {oyetoyan2018}
\bibfield{author}{\bibinfo{person}{Tosin~Daniel Oyetoyan},
  \bibinfo{person}{Bisera Milosheska}, \bibinfo{person}{Mari Grini}, {and}
  \bibinfo{person}{Daniela~Soares Cruzes}.} \bibinfo{year}{2018}\natexlab{}.
\newblock \showarticletitle{Myths and facts about static application security
  testing tools: an action research at telenor digital}. In
  \bibinfo{booktitle}{\emph{International Conference on Agile Software
  Development}}. Springer, Cham, \bibinfo{pages}{86--103}.
\newblock


\bibitem[\protect\citeauthoryear{Paul, Turzo, and Bosu}{Paul
  et~al\mbox{.}}{2021}]%
        {paul2021}
\bibfield{author}{\bibinfo{person}{Rajshakhar Paul},
  \bibinfo{person}{Asif~Kamal Turzo}, {and} \bibinfo{person}{Amiangshu Bosu}.}
  \bibinfo{year}{2021}\natexlab{}.
\newblock \showarticletitle{Why Security Defects Go Unnoticed during Code
  Reviews? A Case-Control Study of the Chromium OS Project}. In
  \bibinfo{booktitle}{\emph{2021 43rd International Conference on Software
  Engineering (ICSE)}}. IEEE.
\newblock


\bibitem[\protect\citeauthoryear{Pereira, Campos, and Vieira}{Pereira
  et~al\mbox{.}}{2019}]%
        {pereira2019}
\bibfield{author}{\bibinfo{person}{Jose~D'Abruzzo Pereira},
  \bibinfo{person}{Jo{\~a}o~R Campos}, {and} \bibinfo{person}{Marco Vieira}.}
  \bibinfo{year}{2019}\natexlab{}.
\newblock \showarticletitle{An exploratory study on machine learning to combine
  security vulnerability alerts from static analysis tools}. In
  \bibinfo{booktitle}{\emph{2019 9th Latin-American Symposium on Dependable
  Computing (LADC)}}. IEEE, \bibinfo{pages}{1--10}.
\newblock


\bibitem[\protect\citeauthoryear{Perl, Dechand, Smith, Arp, Yamaguchi, Rieck,
  Fahl, and Acar}{Perl et~al\mbox{.}}{2015}]%
        {perl2015}
\bibfield{author}{\bibinfo{person}{Henning Perl}, \bibinfo{person}{Sergej
  Dechand}, \bibinfo{person}{Matthew Smith}, \bibinfo{person}{Daniel Arp},
  \bibinfo{person}{Fabian Yamaguchi}, \bibinfo{person}{Konrad Rieck},
  \bibinfo{person}{Sascha Fahl}, {and} \bibinfo{person}{Yasemin Acar}.}
  \bibinfo{year}{2015}\natexlab{}.
\newblock \showarticletitle{Vccfinder: Finding potential vulnerabilities in
  open-source projects to assist code audits}. In
  \bibinfo{booktitle}{\emph{Proceedings of the 22nd ACM SIGSAC Conference on
  Computer and Communications Security}}. \bibinfo{pages}{426--437}.
\newblock


\bibitem[\protect\citeauthoryear{Poel}{Poel}{2010}]%
        {poel2010}
\bibfield{author}{\bibinfo{person}{Nico Poel}.}
  \bibinfo{year}{2010}\natexlab{}.
\newblock \emph{\bibinfo{title}{Automated Security Review of PHP Web
  Applications with Static Code Analysis}}.
\newblock \bibinfo{thesistype}{Master's\ thesis}. \bibinfo{school}{University
  of Groningen}.
\newblock


\bibitem[\protect\citeauthoryear{Pornprasit and Tantithamthavorn}{Pornprasit
  and Tantithamthavorn}{2021}]%
        {pornprasit2021}
\bibfield{author}{\bibinfo{person}{Chanathip Pornprasit} {and}
  \bibinfo{person}{Chakkrit Tantithamthavorn}.}
  \bibinfo{year}{2021}\natexlab{}.
\newblock \showarticletitle{JITLine: A Simpler, Better, Faster, Finer-grained
  Just-In-Time Defect Prediction}.
\newblock \bibinfo{journal}{\emph{arXiv preprint arXiv:2103.07068}}
  (\bibinfo{year}{2021}).
\newblock


\bibitem[\protect\citeauthoryear{Rahman, Khatri, Barr, and Devanbu}{Rahman
  et~al\mbox{.}}{2014}]%
        {rahman2014}
\bibfield{author}{\bibinfo{person}{Foyzur Rahman}, \bibinfo{person}{Sameer
  Khatri}, \bibinfo{person}{Earl~T Barr}, {and} \bibinfo{person}{Premkumar
  Devanbu}.} \bibinfo{year}{2014}\natexlab{}.
\newblock \showarticletitle{Comparing static bug finders and statistical
  prediction}. In \bibinfo{booktitle}{\emph{Proceedings of the 36th
  International Conference on Software Engineering}}.
  \bibinfo{pages}{424--434}.
\newblock


\bibitem[\protect\citeauthoryear{Ribeiro, Meirelles, Lago, and Kon}{Ribeiro
  et~al\mbox{.}}{2019}]%
        {ribeiro2019}
\bibfield{author}{\bibinfo{person}{Athos Ribeiro}, \bibinfo{person}{Paulo
  Meirelles}, \bibinfo{person}{Nelson Lago}, {and} \bibinfo{person}{Fabio
  Kon}.} \bibinfo{year}{2019}\natexlab{}.
\newblock \showarticletitle{Ranking warnings from multiple source code static
  analyzers via ensemble learning}. In \bibinfo{booktitle}{\emph{Proceedings of
  the 15th International Symposium on Open Collaboration}}.
  \bibinfo{pages}{1--10}.
\newblock


\bibitem[\protect\citeauthoryear{Sahal and Tosun}{Sahal and Tosun}{2018}]%
        {sahal2018}
\bibfield{author}{\bibinfo{person}{Emre Sahal} {and} \bibinfo{person}{Ayse
  Tosun}.} \bibinfo{year}{2018}\natexlab{}.
\newblock \showarticletitle{Identifying bug-inducing changes for code
  additions}. In \bibinfo{booktitle}{\emph{Proceedings of the 12th ACM/IEEE
  International Symposium on Empirical Software Engineering and Measurement}}.
  \bibinfo{pages}{1--2}.
\newblock


\bibitem[\protect\citeauthoryear{Scandariato, Walden, Hovsepyan, and
  Joosen}{Scandariato et~al\mbox{.}}{2014}]%
        {scandariato2014}
\bibfield{author}{\bibinfo{person}{Riccardo Scandariato},
  \bibinfo{person}{James Walden}, \bibinfo{person}{Aram Hovsepyan}, {and}
  \bibinfo{person}{Wouter Joosen}.} \bibinfo{year}{2014}\natexlab{}.
\newblock \showarticletitle{Predicting vulnerable software components via text
  mining}.
\newblock \bibinfo{journal}{\emph{IEEE Transactions on Software Engineering}}
  \bibinfo{volume}{40}, \bibinfo{number}{10} (\bibinfo{year}{2014}),
  \bibinfo{pages}{993--1006}.
\newblock


\bibitem[\protect\citeauthoryear{Seacord}{Seacord}{2005}]%
        {seacord2005}
\bibfield{author}{\bibinfo{person}{Robert~C Seacord}.}
  \bibinfo{year}{2005}\natexlab{}.
\newblock \bibinfo{booktitle}{\emph{Secure Coding in C and C++}}.
\newblock \bibinfo{publisher}{Pearson Education}.
\newblock


\bibitem[\protect\citeauthoryear{Shahriar and Zulkernine}{Shahriar and
  Zulkernine}{2012}]%
        {shahriar2012}
\bibfield{author}{\bibinfo{person}{Hossain Shahriar} {and}
  \bibinfo{person}{Mohammad Zulkernine}.} \bibinfo{year}{2012}\natexlab{}.
\newblock \showarticletitle{Mitigating program security vulnerabilities:
  Approaches and challenges}.
\newblock \bibinfo{journal}{\emph{ACM Computing Surveys (CSUR)}}
  \bibinfo{volume}{44}, \bibinfo{number}{3} (\bibinfo{year}{2012}),
  \bibinfo{pages}{1--46}.
\newblock


\bibitem[\protect\citeauthoryear{Shin, Meneely, Williams, and Osborne}{Shin
  et~al\mbox{.}}{2010}]%
        {shin2010}
\bibfield{author}{\bibinfo{person}{Yonghee Shin}, \bibinfo{person}{Andrew
  Meneely}, \bibinfo{person}{Laurie Williams}, {and} \bibinfo{person}{Jason~A
  Osborne}.} \bibinfo{year}{2010}\natexlab{}.
\newblock \showarticletitle{Evaluating complexity, code churn, and developer
  activity metrics as indicators of software vulnerabilities}.
\newblock \bibinfo{journal}{\emph{IEEE transactions on software engineering}}
  \bibinfo{volume}{37}, \bibinfo{number}{6} (\bibinfo{year}{2010}),
  \bibinfo{pages}{772--787}.
\newblock


\bibitem[\protect\citeauthoryear{Shin and Williams}{Shin and Williams}{2013}]%
        {shin2013}
\bibfield{author}{\bibinfo{person}{Yonghee Shin} {and} \bibinfo{person}{Laurie
  Williams}.} \bibinfo{year}{2013}\natexlab{}.
\newblock \showarticletitle{Can traditional fault prediction models be used for
  vulnerability prediction?}
\newblock \bibinfo{journal}{\emph{Empirical Software Engineering}}
  \bibinfo{volume}{18}, \bibinfo{number}{1} (\bibinfo{year}{2013}),
  \bibinfo{pages}{25--59}.
\newblock


\bibitem[\protect\citeauthoryear{Smith, Johnson, Murphy-Hill, Chu, and
  Lipford}{Smith et~al\mbox{.}}{2018}]%
        {smith2018}
\bibfield{author}{\bibinfo{person}{Justin Smith}, \bibinfo{person}{Brittany
  Johnson}, \bibinfo{person}{Emerson Murphy-Hill}, \bibinfo{person}{Bill Chu},
  {and} \bibinfo{person}{Heather~Richter Lipford}.}
  \bibinfo{year}{2018}\natexlab{}.
\newblock \showarticletitle{How developers diagnose potential security
  vulnerabilities with a static analysis tool}.
\newblock \bibinfo{journal}{\emph{IEEE Transactions on Software Engineering}}
  \bibinfo{volume}{45}, \bibinfo{number}{9} (\bibinfo{year}{2018}),
  \bibinfo{pages}{877--897}.
\newblock


\bibitem[\protect\citeauthoryear{Smyth}{Smyth}{2017}]%
        {smyth2017}
\bibfield{author}{\bibinfo{person}{Vincent Smyth}.}
  \bibinfo{year}{2017}\natexlab{}.
\newblock \showarticletitle{Software vulnerability management: how intelligence
  helps reduce the risk}.
\newblock \bibinfo{journal}{\emph{Network Security}} \bibinfo{volume}{2017},
  \bibinfo{number}{3} (\bibinfo{year}{2017}), \bibinfo{pages}{10--12}.
\newblock


\bibitem[\protect\citeauthoryear{Tantithamthavorn, Hassan, and
  Matsumoto}{Tantithamthavorn et~al\mbox{.}}{2018}]%
        {tantithamthavorn2018}
\bibfield{author}{\bibinfo{person}{Chakkrit Tantithamthavorn},
  \bibinfo{person}{Ahmed~E Hassan}, {and} \bibinfo{person}{Kenichi Matsumoto}.}
  \bibinfo{year}{2018}\natexlab{}.
\newblock \showarticletitle{The impact of class rebalancing techniques on the
  performance and interpretation of defect prediction models}.
\newblock \bibinfo{journal}{\emph{IEEE Transactions on Software Engineering}}
  \bibinfo{volume}{46}, \bibinfo{number}{11} (\bibinfo{year}{2018}),
  \bibinfo{pages}{1200--1219}.
\newblock


\bibitem[\protect\citeauthoryear{Tantithamthavorn, Jiarpakdee, and
  Grundy}{Tantithamthavorn et~al\mbox{.}}{2020}]%
        {tantithamthavorn2020}
\bibfield{author}{\bibinfo{person}{Chakkrit Tantithamthavorn},
  \bibinfo{person}{Jirayus Jiarpakdee}, {and} \bibinfo{person}{John Grundy}.}
  \bibinfo{year}{2020}\natexlab{}.
\newblock \showarticletitle{Explainable AI for Software Engineering}.
\newblock \bibinfo{journal}{\emph{arXiv preprint arXiv:2012.01614}}
  (\bibinfo{year}{2020}).
\newblock


\bibitem[\protect\citeauthoryear{Tantithamthavorn, McIntosh, Hassan, and
  Matsumoto}{Tantithamthavorn et~al\mbox{.}}{2016}]%
        {tantithamthavorn2016}
\bibfield{author}{\bibinfo{person}{Chakkrit Tantithamthavorn},
  \bibinfo{person}{Shane McIntosh}, \bibinfo{person}{Ahmed~E Hassan}, {and}
  \bibinfo{person}{Kenichi Matsumoto}.} \bibinfo{year}{2016}\natexlab{}.
\newblock \showarticletitle{An empirical comparison of model validation
  techniques for defect prediction models}.
\newblock \bibinfo{journal}{\emph{IEEE Transactions on Software Engineering}}
  \bibinfo{volume}{43}, \bibinfo{number}{1} (\bibinfo{year}{2016}),
  \bibinfo{pages}{1--18}.
\newblock


\bibitem[\protect\citeauthoryear{TIOBE}{TIOBE}{[n.d.]}]%
        {TIOBE}
\bibfield{author}{\bibinfo{person}{TIOBE}.} \bibinfo{year}{[n.d.]}\natexlab{}.
\newblock \bibinfo{title}{TIOBE Index}.
\newblock
\newblock
\urldef\tempurl%
\url{https://www.tiobe.com/tiobe-index/}
\showURL{%
\tempurl}


\bibitem[\protect\citeauthoryear{Viega, Bloch, Kohno, and McGraw}{Viega
  et~al\mbox{.}}{2002}]%
        {viega2002}
\bibfield{author}{\bibinfo{person}{John Viega}, \bibinfo{person}{JT Bloch},
  \bibinfo{person}{Tadayoshi Kohno}, {and} \bibinfo{person}{Gary McGraw}.}
  \bibinfo{year}{2002}\natexlab{}.
\newblock \showarticletitle{Token-based scanning of source code for security
  problems}.
\newblock \bibinfo{journal}{\emph{ACM Transactions on Information and System
  Security (TISSEC)}} \bibinfo{volume}{5}, \bibinfo{number}{3}
  (\bibinfo{year}{2002}), \bibinfo{pages}{238--261}.
\newblock


\bibitem[\protect\citeauthoryear{Wagner and Sametinger}{Wagner and
  Sametinger}{2014}]%
        {wagner2014}
\bibfield{author}{\bibinfo{person}{Andreas Wagner} {and}
  \bibinfo{person}{Johannes Sametinger}.} \bibinfo{year}{2014}\natexlab{}.
\newblock \showarticletitle{Using the Juliet test suite to compare static
  security scanners}. In \bibinfo{booktitle}{\emph{2014 11th International
  Conference on Security and Cryptography (SECRYPT)}}. IEEE,
  \bibinfo{pages}{1--9}.
\newblock


\bibitem[\protect\citeauthoryear{Walden, Stuckman, and Scandariato}{Walden
  et~al\mbox{.}}{2014}]%
        {walden2014}
\bibfield{author}{\bibinfo{person}{James Walden}, \bibinfo{person}{Jeff
  Stuckman}, {and} \bibinfo{person}{Riccardo Scandariato}.}
  \bibinfo{year}{2014}\natexlab{}.
\newblock \showarticletitle{Predicting vulnerable components: Software metrics
  vs text mining}. In \bibinfo{booktitle}{\emph{2014 IEEE 25th international
  symposium on software reliability engineering}}. IEEE,
  \bibinfo{pages}{23--33}.
\newblock


\bibitem[\protect\citeauthoryear{Wheeler}{Wheeler}{[n.d.]}]%
        {flawfinder}
\bibfield{author}{\bibinfo{person}{David Wheeler}.}
  \bibinfo{year}{[n.d.]}\natexlab{}.
\newblock \bibinfo{title}{Flawfinder}.
\newblock
\newblock
\urldef\tempurl%
\url{https://dwheeler.com/flawfinder/}
\showURL{%
\tempurl}


\bibitem[\protect\citeauthoryear{Wilcoxon}{Wilcoxon}{1992}]%
        {wilcoxon1992}
\bibfield{author}{\bibinfo{person}{Frank Wilcoxon}.}
  \bibinfo{year}{1992}\natexlab{}.
\newblock \showarticletitle{Individual comparisons by ranking methods}.
\newblock In \bibinfo{booktitle}{\emph{Breakthroughs in statistics}}.
  \bibinfo{publisher}{Springer}, \bibinfo{pages}{196--202}.
\newblock


\bibitem[\protect\citeauthoryear{Xie, Chou, and Engler}{Xie
  et~al\mbox{.}}{2003}]%
        {xie2003}
\bibfield{author}{\bibinfo{person}{Yichen Xie}, \bibinfo{person}{Andy Chou},
  {and} \bibinfo{person}{Dawson Engler}.} \bibinfo{year}{2003}\natexlab{}.
\newblock \showarticletitle{Archer: using symbolic, path-sensitive analysis to
  detect memory access errors}. In \bibinfo{booktitle}{\emph{Proceedings of the
  9th European software engineering conference held jointly with 11th ACM
  SIGSOFT international symposium on Foundations of software engineering}}.
  \bibinfo{pages}{327--336}.
\newblock


\bibitem[\protect\citeauthoryear{Yang, Xia, Lo, Bi, Grundy, and Yang}{Yang
  et~al\mbox{.}}{2020}]%
        {yang2020}
\bibfield{author}{\bibinfo{person}{Yanming Yang}, \bibinfo{person}{Xin Xia},
  \bibinfo{person}{David Lo}, \bibinfo{person}{Tingting Bi},
  \bibinfo{person}{John Grundy}, {and} \bibinfo{person}{Xiaohu Yang}.}
  \bibinfo{year}{2020}\natexlab{}.
\newblock \showarticletitle{Predictive Models in Software Engineering:
  Challenges and Opportunities}.
\newblock \bibinfo{journal}{\emph{arXiv preprint arXiv:2008.03656}}
  (\bibinfo{year}{2020}).
\newblock


\bibitem[\protect\citeauthoryear{Yoon, Jin, and Jung}{Yoon
  et~al\mbox{.}}{2014}]%
        {yoon2014}
\bibfield{author}{\bibinfo{person}{Jongwon Yoon}, \bibinfo{person}{Minsik Jin},
  {and} \bibinfo{person}{Yungbum Jung}.} \bibinfo{year}{2014}\natexlab{}.
\newblock \showarticletitle{Reducing false alarms from an industrial-strength
  static analyzer by SVM}. In \bibinfo{booktitle}{\emph{2014 21st Asia-Pacific
  Software Engineering Conference}}, Vol.~\bibinfo{volume}{2}. IEEE,
  \bibinfo{pages}{3--6}.
\newblock


\bibitem[\protect\citeauthoryear{Zeng, Lin, Pan, Tai, and Zhang}{Zeng
  et~al\mbox{.}}{2020}]%
        {zeng2020}
\bibfield{author}{\bibinfo{person}{Peng Zeng}, \bibinfo{person}{Guanjun Lin},
  \bibinfo{person}{Lei Pan}, \bibinfo{person}{Yonghang Tai}, {and}
  \bibinfo{person}{Jun Zhang}.} \bibinfo{year}{2020}\natexlab{}.
\newblock \showarticletitle{Software Vulnerability Analysis and Discovery using
  Deep Learning Techniques: A Survey}.
\newblock \bibinfo{journal}{\emph{IEEE Access}} (\bibinfo{year}{2020}).
\newblock


\bibitem[\protect\citeauthoryear{Zimmermann, Nagappan, and Williams}{Zimmermann
  et~al\mbox{.}}{2010}]%
        {zimmermann2010}
\bibfield{author}{\bibinfo{person}{Thomas Zimmermann},
  \bibinfo{person}{Nachiappan Nagappan}, {and} \bibinfo{person}{Laurie
  Williams}.} \bibinfo{year}{2010}\natexlab{}.
\newblock \showarticletitle{Searching for a needle in a haystack: Predicting
  security vulnerabilities for windows vista}. In
  \bibinfo{booktitle}{\emph{2010 Third International Conference on Software
  Testing, Verification and Validation}}. IEEE, \bibinfo{pages}{421--428}.
\newblock


\end{thebibliography}

\end{document}